\documentclass[11pt]{article}
\pdfoutput=1
\usepackage{jheppub}
\usepackage{amssymb,amsfonts}
\usepackage{mathrsfs}
\usepackage{epsfig}
\let\savenumberline\numberline
\def\numberline#1{\savenumberline{#1.}}
\makeatletter
\renewcommand{\@seccntformat}[1]{\csname the#1\endcsname.\,\,}
\makeatother

\newcommand{\R}{{\bf R}}

\newcommand{\CD}{{\cal D}}
\newcommand{\CE}{{\cal E}}
\newcommand{\CF}{{\cal F}}
\newcommand{\CG}{{\cal G}}

\newcommand{\CI}{{\cal I}}

\newcommand{\CL}{{\cal L}}
\newcommand{\CM}{{\cal M}}
\newcommand{\CN}{{\cal N}}
\newcommand{\CO}{{\cal O}}

\newcommand{\CV}{{\cal V}}
\newcommand{\CW}{{\cal W}}

\newcommand{\CZ}{{\cal Z}}

\newcommand{\RD}{{{\mathrm D}}}
\newcommand{\barRD}{{\bar{\mathrm D}}}
\newcommand{\SD}{{\mathscr D}}
\newcommand{\SF}{{\mathscr F}}
\newcommand{\SM}{{\mathscr M}}

\newcommand{\gh}{{\textsf{gh}}}

\newcommand{\p}{\partial}
\renewcommand{\bar}[1]{\overline{#1}}
\renewcommand{\tilde}[1]{\widetilde{#1}}
\newcommand{\be}{\begin{equation}}
\newcommand{\ee}{\end{equation}}
\newcommand{\bea}{\begin{eqnarray}}
\newcommand{\eea}{\end{eqnarray}}

\newcommand{\ie}{{\it i.e.}}
\newcommand{\eg}{{\it e.g.}}
\newcommand{\diff}{{\rm Diff}}

\makeatletter
\def\@fpheader{\relax}
\makeatother
\usepackage{graphicx}
\usepackage{latexsym}

\title{\ \vspace{1.5in} \\ Topological Quantum Gravity of the Ricci Flow}
\author{Alexander Frenkel${}^a$, Petr Ho\v{r}ava${}^{b,c}$ and Stephen Randall${}^{b,c}$}
\affiliation{${}^a$Stanford Institute for Theoretical Physics and Department of Physics\\
Stanford University, Stanford, CA, 94305-4060, USA\medskip\\
${}^b$Berkeley Center for Theoretical Physics and Department of Physics\\
University of California, Berkeley, CA, 94720-7300, USA\medskip\\
${}^c$Physics Division, Lawrence Berkeley National Laboratory\\
Berkeley, CA 94720-8162, USA}
\abstract{We present a family of topological quantum gravity theories associated with the geometric theory of the Ricci flow on Riemannian manifolds.  First we use BRST quantization to construct a ``primitive'' topological Lifshitz-type theory for only the spatial metric, with spatial diffeomorphism invariance and no gauge symmetry, associated with Hamilton's Ricci flow:  Hamilton's flow equation appears as the localization equation of the primitive theory.  Then we extend the primitive theory by gauging foliation-preserving spacetime symmetries.  Crucially, all our theories are required to exhibit an ${\cal N}=2$ extended BRST symmetry.  First, we gauge spatial diffeomorphisms, and show that this gives us access to the mathematical technique known as the DeTurck trick.  Finally, we gauge foliation-preserving time reparametrizations, both with the projectable and nonprojectable lapse function.    The path integral of the full theory is localized to the solutions of Ricci-type flow equations, generalizing those of Perelman.  The role of Perelman's dilaton is played by the nonprojectable lapse function.  Perelman's ${\cal F}$-functional appears as the superpotential of our theory.  Since there is no spin-statistics theorem in nonrelativistic quantum field theory, the two supercharges of our gravity theory do not have to be interpreted as BRST charges and, after the continuation to real time, the theory can be studied as a candidate for nonrelativistic quantum gravity with propagating bosonic and fermionic degrees of freedom.}
\begin{document}
\maketitle
\section{Introduction}

In this paper, we bring together three distinct and previously rather unrelated subjects: The geometry of the Ricci-type flows on Riemannian manifolds, topological quantum field theory, and nonrelativistic Lifshitz-type quantum gravity.

The Ricci flow on Riemannian manifolds, governed by the Ricci flow equation
\be
\frac{\p g_{ij}}{\p t}=-2R_{ij},
\label{eehamrf}
\ee
was introduced by Richard Hamilton in 1982 \cite{hamfirst}, as a potentially powerful tool for addressing some of the deep open questions in differential geometry and topology of low-dimensional manifolds.  This program has been -- and continues to be -- very successful, leading to Grisha Perelman's celebrated proof \cite{perel1,perel2,perel3} of the Poincar\'e conjecture, the proof of Thurston's geometrization conjecture for 3-manifolds, a new independent proof of the uniformization theorem for 2-manifolds \cite{tiann}, and more recently the proof of the generalized Smale conjecture \cite{smale,bamkl1,bamkl2,kl1,kl2}.  One of the important stepping-stones was Perelman's addition of a ``dilaton'' field $\phi$ to the spatial metric, and his formulation of the combined flow equations of $g_{ij}$ and $\phi$ as a gradient flow for the so-called $\CF$-functional,
\be
\CF(g_{ij},\phi)=2\int d^D x\,e^{-\phi}\sqrt{g}\left\{R+\,g^{ij}\,\p_i\phi\,\p_j\phi\right\}.
\label{eeperlf}
\ee
In the process of proving the consequences of this flow, a truly impressive wealth of many geometric and topological results and insights has been accumulated in the past two decades, with many intriguing questions still remaining open and vigorous investigations being actively pursued.  A comprehensive multi-volume introduction to the mathematics of Ricci flow can be found in \cite{rfi,rf1,rf2,rf3,rf4}.%
\footnote{A comment about our list of references:  Each of the three subjects that we connect in this paper has a hugely extensive literature.  Hence, our list of references is inevitably far from exhaustive; we focus on a relatively short list of papers and books that we find particularly relevant to our construction, plus a longer list of various illuminating reviews.}
Many excellent and mutually complementary mathematical reviews and surveys exist: \cite{hamsur,kleinl,morgant1,morgantb,morgant2,morgant3,topping,hrf,muller,tao}.  Many of the foundational papers (including almost all of Hamilton's papers on the subject prior to 2002 and his influential 1995 survey \cite{hamsur}) are collected in \cite{collrf}.  

\begin{figure}[t!]
  \centering
    \includegraphics[width=0.6\textwidth]{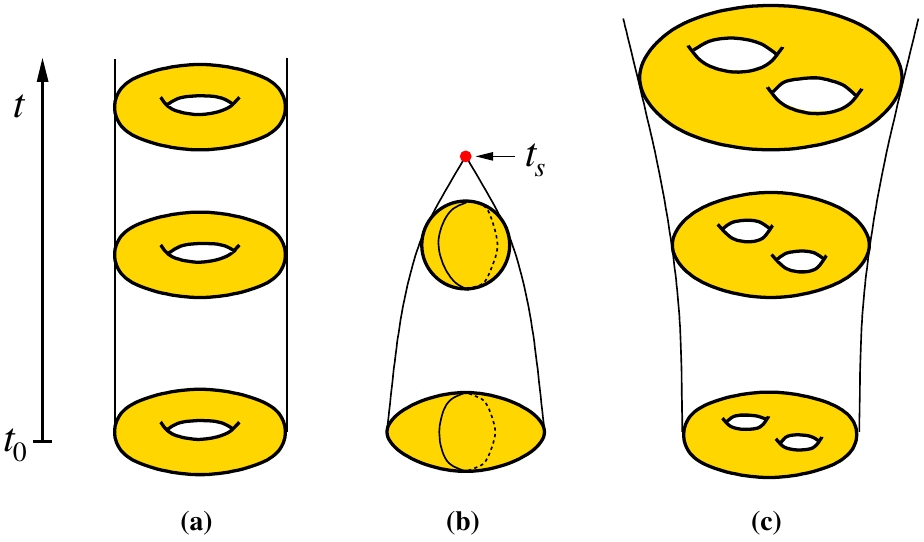}
    \caption{Simple illustrations of the typical behavior of the Ricci flow (\ref{eehamrf}) in $3+1$ dimensions. \textbf{(a):} A Ricci-flat manifold stays constant with time. \textbf{(b):} A manifold with positive sectional curvatures, such as a slightly deformed sphere with bounded spatial inhomogeneities, will round itself out with time and uniformly collapse into a singularity at a finite instant $t_s$. \textbf{(c):} A hyperbolic manifold, with negative sectional curvatures, will expand forever.}
    \label{ffone}
\end{figure}

Topological quantum field theories (of the ``cohomological'' type relevant for this paper) were introduced by Edward Witten in 1988: The first examples included topological Yang-Mills gauge theory \cite{ewtym} in $3+1$ dimensions, topological nonlinear sigma models \cite{ewtsm} in $1+1$ dimensions which later became central in the construction of topological string theory, and the first version of topological gravity \cite{ewtg}.  The central role in the construction is played by the BRST quantization and BRST cohomology.%
\footnote{For the general overview of BRST symmetry, see for example \cite{htbook}.}
An accessible introduction to the general concept of topological quantum field theories of this cohomological type is in \cite{ewcoho}.  Roughly, for any ``interesting'' differential equation, one can attempt to construct a topological quantum field theory of the cohomological type, whose path integral is expected to localize to the moduli space of the appropriate solutions of the equation.  Ref.~\cite{ewcoho} provides if not an algorithm, then at least an itinerary how to do this.  In this way, topological Yang-Mills theory is associated with the self-duality equation for the field strength of the Yang-Mills connection, and the instanton moduli space.  Physical observables are related to Donaldson invariants of 4-manifolds.  Similarly, the topological sigma model is associated with Gromov's pseudoholomorphic curve equation which describes worldsheet instantons in string theory.  Observables lead to Gromov-Witten invariants.  

In this paper, our main goal is to to construct a topological quantum field theory associated with a generalized family of Ricci flow equations.  The proper setting for this construction is in nonrelativistic quantum gravity, and its supersymmetric and topological generalizations.  Nonrelativistic quantum gravity with anisotropic scaling (in the literature often referred to as Ho\v rava-Lifshitz gravity; we will refer to it in this paper as Lifshitz-type gravity) was introduced in \cite{mqc,lif,grx}.  It has been broadly studied as an example of quantum gravity with improved short-distance behavior, which can explain the numerical lattice results of the Causal Dynamical Triangulations approach to quantum gravity \cite{cdt1,cdt2,cdt3}, and even be power-counting renormalizable in appropriate dimensions; as a tool for nonrelativistic holography, where it leads to a broader set of holographic duals of nonrelativistic systems than bulk relativistic gravity; and for cosmology \cite{shinji}.

\begin{figure}[b!]
  \centering
    \includegraphics[width=0.73\textwidth]{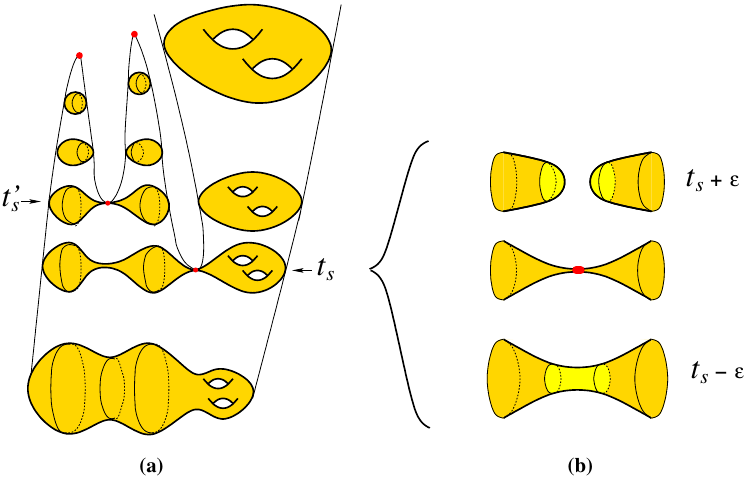}
    \caption{\textbf{(a):} Another illustration of the Ricci flow (\ref{eehamrf}) in $3+1$ dimensions, now involving not only examples of the extinction singularity of the positively curved regions, but also two examples of a generic ``neckpinch'' singularity in finite time (here at $t_s$ and $t_s'$).  \textbf{(b):} The spatial topology change caused by the neckpinch singularity is handled by the geometrical technique of surgery on manifolds \cite{ctc}.  Zooming in on a small vicinity of the singularity, we find the spatial topology of $I\times S^2$ at time $t_s-\varepsilon$.  Surgery replaces it with the union of two 3-balls $B_3\cup B_3$ at $t_s+\varepsilon$, and restarts the Ricci flow.}
    \label{fftwo}
\end{figure}

The mathematical theory of the Ricci flow has been previously connected to physics in several ways.  The relation to the renormalization group flow of nonlinear sigma models in two relativistic dimensions was already stressed and utilized by Perelman in \cite{perel1}; for further developments of this connection, see \cite{tseytlin}.  Another useful connection has been made to numerical general relativity \cite{headw}.  In this paper, we find a new connection between Ricci flows and physics:  We construct a topological quantum field theory of the cohomological type, whose path integral localizes to the solutions of a family of Ricci flow equations.  This theory will inevitably take the form of a topological norelativistic quantum gravity. That such a topological theory of Lifshitz-type gravity associated with the evolution equations of the Ricci type should exist was first conjectured during the work on \cite{mqc}, see also the discussion in \S 1.3 of \cite{gen}.  The purpose of this paper is to fill this gap, and to present an explicit construction which links the mathematical theory of the Ricci flow to the physics of topological quantum field theory and quantum gravity.

This paper is organized as follows.  We build our topological quantum gravity of the Ricci flow in stages, introducing a simplest version of nonrelativistic topological gravity first, and then bringing in additional steps and features needed to make contact with the Perelman theory of the Ricci flow.

In Section~\ref{ssprim}, we construct a ``primitive'' theory of topological nonrelativistic quantum gravity.  The dynamical field in the primitive theory is the spatial metric $g_{ij}(t,x^k)$ on a $D+1$ dimensional spacetime, which carries a natural foliation structure by $D$-dimensional leaves $\Sigma$ of constant time $t$.  It is true that $D=3$ appears to be the most immediately interesting case, both in physics and in mathematics, but our construction is more general than that, so we present it in $D$ dimensions.  The symmetries are all local topological deformations of $g_{ij}$.  In addition to the topological BRST charge, we require the existence of an anti-BRST supercharge $\bar Q$, and construct the gauge-fixed primitive theory in an appropriately defined $\CN=2$ superspace.  This theory is particularly interesting when the dynamical exponent $z$ (which is a measure of the anisotropy between time and space) is equal to two:  While it is well-known that Hamilton's Ricci flow equation (\ref{eehamrf}) cannot be derived from a variational principle, we find that (\ref{eehamrf}) \textit{represents the locatization equation} in our primitive theory, for certain values of the coupling constants.

Much of modern theoretical physics is built around the concept of gauge symmetries.  In the context of quantum gravity, it is natural to expect some form of spacetime diffeomorphism symmetry.  The primitive theory constructed in Section~\ref{ssprim} has no spacetime gauge symmetries: It is only invariant under time-independent spatial diffeomorphisms.  In Section~\ref{ssdiffs}, we take the first step to remedy this, and we gauge spatial diffeomorphisms.  This is again done in $\CN=2$ superspace, by introducing the shift vector $n^i(t,x^j)$ and its superpartners.  It is in this theory with spatial diffeomorphisms promoted to a gauge symmetry where we find a natural setting for an important Ricci-flow technique known in the mathematical literature as the ``DeTurck trick.''  It simply appears via possible choices of gauge fixing conditions.

In Section~\ref{sstime}, we extend the gauge symmetry to include time reparametrizations, and thus promote the symmetries from spacetime-dependent spatial diffeomorphisms of Section~\ref{ssdiffs} to the full gauge symmetry generally expected in Lifshitz-type quantum gravity: Foliation-preserving diffeomorphisms of spacetime.  The gauging is accomplished by introducing the lapse function $n$ and its $\CN=2$ superpartners, which can be either projectable (\ie , dependent only on time), or nonprojectable, $n(t,x^i)$.  We concentrate on the nonprojectable version of the theory, and reach two conclusions, which represent the central results of this paper:  (1) the role of Perelman's ``dilaton'' is played in our theory by the lapse function (more precisely, $\phi=-\log n$), and (2) Perelman's $\CF$-functional arises simply as the $\CN=2$ superpotential in our topological gravity.

Our construction leads to a multi-parameter family of topological quantum gravities, whose localization equations represent a multi-parameter generalization of Perelman's Ricci flow equations for the fields $g_{ij}$ and $\phi$, parametrized by the values of general couplings in our topological gravity Lagrangian with $z=2$ dynamical scaling.  We list some open questions and challenges in Section~\ref{ssc}.

\section{The primitive theory}
\label{ssprim}

In Lifshitz-type gravity, one can describe the dynamics of spacetime geometry using the fields of the ADM formalism, first developed in the Hamiltonian description of general relativity \cite{adm}.  These ADM variables consist of the spatial metric $g_{ij}$, the shift vector $n^i$, and the lapse function $n$,%
\footnote{The lapse and shift variables are usually denoted in the literature by the capital letters $N$ and $N^i$.  In this paper, we reserve $N$ and $N^i$ to denote the superfields whose lowest components are the lapse and shift $n$ and $n^i$.}
and they were originally viewed as a decomposition of the full relativistic spacetime metric.  In Lifshitz-type gravity, these fields define \textit{two distinct} geometric length elements $d\tau$ and $d\sigma$ on spacetime,
\be
d\tau=n\,dt,\qquad d\sigma^2=g_{ij}(dx^i+n^idt)(dx^j+n^jdt).
\label{eedtau}
\ee
\textit{A priori}, these two elements are unrelated, and the distances they define are measured in two different units:  The spatial length scale $L$ and the time scale $T$.  In such a theory with two separate scales, the scaling properties of fields and their derivatives undergo the appropriate refinement in comparison to their relativistic counterparts.  In the traditional way of assigning classical dimensions to the building blocks of Lifshitz gravity, one assigns the coordinate element $dx^i$ the dimension of length, $[dx^i]=L$, and the time element the dimension of time, $[dt]=T$.  Since the physical distances $d\sigma$ and $d\tau$ also have those same dimensions, $[d\sigma]=L$ and $[d\tau]=T$, we see from (\ref{eedtau}) that $g_{ij}$ and $n$ are both dimensionless, and that the dimension of the shift vector is $[n^i]=L/T$.

In general relativity, the speed of light $c$ is a dimensionful constant of nature which relates space and time distances to each other in a canonical way, and $d\tau$ and $d\sigma$ combine to form the unique spacetime metric, which transforms covariantly under the symmetries of general relativity.  We can naturally set $c=1$ for convenience, which canonically relates $L=T$, and the spacetime metric is then
\be
ds^2_{\mathrm{GR}}=-d\tau^2+d\sigma^2=-(n^2-n_in^i)dt^2+2n_idx^idt+g_{ij}dx^idx^j.
\label{eedsig}
\ee
The theory now has only one scale in which dimensions of fields and their derivatives are measured.

In contrast, in nonrelativistic Lifshitz gravity no such canonical constant of nature $c$ is present, and the two length elements (\ref{eedtau}) and (\ref{eedsig}) cannot be canonically combined into a unique spacetime element.  A relation between the two scales $L$ and $T$ is typically generated by the renormalization-group fixed point appropriate for the system in a particular regime.  Typically, the short-distance physics is dominated by one fixed point, characterized by the relation $T\sim L^z$, where $z$ is the dynamical critical exponent characterizing the short-distance anisotropy between time and space; usually, we have $z>1$.  The long-distance physics is typically governed by another fixed point, usually with $z=1$, resulting from the natural renormalization-group flow of the theory.  Whether the theory is short-distance complete or at least power-counting renormalizable is governed by the value of $z$ at the short-distance fixed point, and the spatial dimension $D$.

\subsection{Preliminaries: the structure of spacetime}

The main purpose of this paper will be to construct an appropriately supersymmetric version of nonrelativistic Lifshitz-type gravity on spacetime manifold $\CM$ of dimension $D+1$, equipped with the further structure of a codimension-one foliation $\CM_\CF$ by spatial slices $\Sigma$ of dimension $D$, which can be thought of as slices of constant time.  There is a natural projection $\pi$ from $\CM$ to the time dimension $\R$, by simply forgetting the location along the leaf $\Sigma$.  We will use coordinates $(t,x^i;i\in{1,\ldots,D})$ on $\CM$, naturally adapted to the foliation $\CM_\CF$ so that $\pi: (t,x^i)\mapsto t$.  

Specific solutions of Ricci-type flow equations often develop interesting singularities, with some simple examples illustrated in Fig.~\ref{ffone} and Fig.~\ref{fftwo}.  Therefore, they may only be defined -- without surgery or some other prescription for continuing through the singularities -- on some open time interval $\CI =(t_0,t_1)\subset\R$, where one or both of $t_0$ and $t_1$ might be finite.  Alternatively, one studies the initial-value problem, on $[t_0,t_1)$ with $t_0$ being the initial time and $t_1$ chosen such that no singularity is encountered for $t<t_1$ in this interval. In this paper, we focus on constructing the description of our quantum gravity theories on such a smooth patch, and leave the fascinating question of the singularities (such as the exctinction singularities of Fig.~\ref{ffone}, or the topology-changing ``neckpinch'' singularities of Fig.~\ref{fftwo}) and their physical interpretation for future study.  Our time manifold $\CM_0$ should therefore be interpreted as either $\R$ when time extends for all eternity, or an open interval $\CI\subset\R$, or the intial-value problem interval $[t_0,t_1)$, as appropriate.  

For simplicity, in this paper we focus on the case of compact $\Sigma$.  We fully expect our theory to describe the noncompact case as well (when $\Sigma$ is a complete Riemannian manifold), but the precise formulation would require a careful discussion of the suitable behavior near the appropriately defined spacetime infinity (in the sense of \cite{aci}), which goes beyond the scope of the work reported here.

In this section, we begin with a simpler task, and construct a more primitive topological gravity theory for the special case when the spacetime manifold is canonically a direct product, $\CM=\Sigma\times\CM_0$, with the time manifold $\CM_0\subset\R$ as explained above.  This construction is simpler because the only dynamical field is the spatial metric $g_{ij}(t,x^k)$ and its superpartners implied by the topological symmetry.  The time dimension is assumed to carry a constant nondynamical metric, and there is no secondary spacetime gauge invariance besides the topological symmetry.  We will refer to this theory as the ``primitive'' theory.

\subsection{Fields and symmetries}

The only dynamical field of the primitive theory will be the spacetime-dependent spatial metric $g_{ij}(t,x^k)$.  (We will use Penrose's ``abstract index'' notation throughout, for all our fields.)  The gauge symmetry will be the topological gauge symmetry, given by all local deformations of the metric,
\be
\label{tgsg}
\delta g_{ij}(t,x^k)=\xi_{ij}(t,x^k).
\ee
We anticipate that due to this very large gauge symmetry, our theory will have no propagating local degrees of freedom (such as gravitons), but it may still have a nontrivial global structure.

The only action that is invariant under the topological gauge symetry (\ref{tgsg}) would be a sum of topological invariants built from the spatial metric, and therefore does not yet define a meaningful path integral.  The path-integral representation of this theory comes entirely from the ``gauge-fixing'' of (\ref{tgsg}) using the BRST method:  One replaces the local gauge symmetry with a global symmetry, generated by a supercharge $Q$ which squares to zero,
\be
Q^2=0.
\ee
This BRST supercharge $Q$ maps $g_{ij}$ to a ghost field $\psi_{ij}(t,x^k)$ which is the section of the same bundle as the gauge transformation parameter $\xi_{ij}$, but carries the opposite (\ie , fermionic) statistics.  Thus, our first BRST multiplet is
\be
Qg_{ij}=\psi_{ij},\qquad Q\psi_{ij}=0.
\ee
The next step is to choose a gauge fixing condition: a local functional $\SF^J$ of $g_{ij}$ and its derivatives, designed such that the path integral of the theory will localize to the space of solutions to $\SF^J=0$.  For judiciously chosen $\SF^J$, the space of such solutions is finite-dimensional, and typically of great geometric interest.  In the process of choosing $\SF^J$, one chooses the bundle on spacetime, and $\SF^J$ will be a section of this bundle.  To implement the gauge fixing and to make sense of the path integral, one then introduces a trivial BRST multiplet consisting of a fermion ``antighost'' $\chi_J$ and the bosonic auxiliary field $B_J$,
\be
Q\chi_{J}=B_J,\qquad Q B_J=0.
\ee
We assign a ``ghost number'' $\gh$: the ghost and antighost are assigned $\gh(\psi_{ij})=1$ and $\gh(\chi_J)=-1$, while $\gh(g_{ij})=\gh(B_J)=0$.  Consequently, the supercharge $Q$ has $\gh(Q)=1$.  Classically, one may start with the requirement that $\gh$ be conserved; quantum mechanically, however, there are often anomalies in this global symmetry, which play an important role in determining the dimensions of the moduli spaces to which the path integral is localized, and what insertions of various observables may be needed to make any correlation function non-vanishing.

With these fields, one then constructs an action
\be
S=\int dt\,d^D x\,\{Q,\Psi\},
\ee
where the ``gauge-fixing fermion'' $\Psi$ is $\sim \chi_J\SF^J$.  We require that $S$ preserve the ghost number symmetry, therefore the gauge-fixing fermion must have $\gh(\Psi)=-1$. States and physical operators in this theory are defined as the cohomology classes of the BRST charge $Q$ on the spaces of all states and operators built from the available fields.

\subsection{Extended BRST superalgebra}

In general, the antighost field does not have to be (and typically indeed is not) the section of the same bundle as the ghost field.  In our topological gravity, we wish to choose as our gauge fixing condition a functional whose vanishing will imply the Ricci-type flow of the metric $g_{ij}$,
\be
\SF^J\sim \frac{\p g_{ij}}{\p t}+2R_{ij}+\ldots.
\ee
Hence, in this case, we have $J\equiv (ij)$, and the ghost and antighost fields \textit{are} sections of the same bundle.

Since the ghost and antighost fields are sections of the same bundle, it is possible to demand that our theory has an additional symmetry, which exchanges the ghosts with antighosts.  Some early examples of topological field theories with this additional ghost-antighost symmetry include the harmonic topological sigma models and the topological rigid string \cite{trst1,trst2} (see also \cite{cmr,trst3}).  The partition function in such theories typically evaluates the appropriately defined Euler number of the moduli space of solutions of the localization equation \cite{blau}.  Topological field theories with the ghost-antighost symmetry later became known as ``balanced theories'' \cite{dmbala}.
We will indeed take advantage of this possibility, and simply {\it postulate\/} that our theory has a second real supercharge $\bar Q$, which also squares to zero,
\be
\label{eeqqzero}
Q^2=0,\qquad \bar Q^2=0.
\ee
We will refer to $\bar Q$ as the ``anti-BRST charge.''  It carries $\gh(\bar Q)=-1$.  Note that as a consequence of the symmetry between the ghosts and antighosts, the global symmetry associated with the ghost number $\gh$ will be non-anomalous.  

In order to complete our superalgebra of supercharges $Q$ and $\bar Q$, we need to decide what their anticommutator should be.  One option would be to simply set it to zero.  Indeed, the anti-BRST charge and the extended BRST algebra was first discovered in the context of gauge-fixing relativistic Yang-Mills gauge theories \cite{antib,ojima}, where $\bar Q$ was found to anticommute with $Q$.  For our purposes it will be crucial to choose another, more interesting possibility consistent with (\ref{eeqqzero}), whereby the supercharges anticommute up to a time translation generator,
\be
\{Q,\bar Q\}=\p_t.
\label{eeqqt}
\ee
This algebra is a natural deformation of the extended BRST anti-BRST algebra found originally in the relativistic setting of Yang-Mills theories \cite{antib,ojima}. In the relativistic case, there simply is no suitable candidate, consistent with Lorentz invariance, for a bosonic symmetry generator that could appear on the right-hand side of (\ref{eeqqt}).  In the nonrelativistic theory, the time translation generator can naturally appear, and our topological gravity will take advantage of this possibility.%
\footnote{The supersymmetric structure bears formal similarity to the supersymmetric treatment of stochastic quantization.  The analogy between Lifshitz-type gravities satisfying the detailed balance condition, and stochastic quantization of a gravity theory in one lower dimension, was pointed out in \cite{mqc,lif}; see also \cite{dijkstoch} and \cite{orlref}.}

Requiring the existence of the second supercharge $\bar Q$ and the extended superalgebra is beneficial for two reasons:  It not only allows us to make it easier to implement the ghost-antighost symmetry but, more importantly, it will also guarantee that the flow equations on which the path integral localizes are {\it gradient\/} flow equations.

\subsection{$\CN=2$ superspace extension of time}

In order to proceed in the most efficient way, it is very natural to organize all component fields into superfields.  Thus, we extend the spacetime manifold into a supermanifold $\SM$ of dimension $D+1|2$, with coordinates $(t,x^i,\theta,\bar\theta)$, where $\theta$ and $\bar\theta$ are two {\it real\/} anticommuting coordinates.%
\footnote{We stress that in this paper, we utilize the oft-used physics convention, in which the bar on top of $\theta$ etc.~is simply an additional index, and never a (complex) conjugation operation.  Thus, $\theta$ and $\bar\theta$ denote two \textit{real} Grassmannian variables, $Q$ and $\bar Q$ are two independent real supercharges, and so on.}
On this supermanifold, we combine the spatial metric, and its ghost, antighost and bosonic auxilary field into the spatial metric superfield
\be
G_{ij}=g_{ij}+\theta\psi_{ij}+\bar\theta\chi_{ij}+\theta\bar\theta B_{ij}.
\ee

Our construction of topological gravity theory involves supersymmetry with two supercharges $Q$ and $\bar Q$.  It will be convenient to formulate the theory directly in the language of superfields and other geometric objects and operations on $\SM$, instead of using the cumbersome component field formulation.  The superspace $\SM$ inherits a natural foliation $\SM_\CF$, again by leaves of the bosonic space $\Sigma$, and therefore is a codimension-$(1|2)$ foliation.  Thus, our bosonic time dimension $\CM_0\subset\R$ is promoted to a supermanifold $\SM_0$ of dimension $(1|2)$, with coordinates $(t,\theta,\bar\theta)$, which we will naturally refer to as ``supersymmetric time'', or ``supertime'' for short.  The projection from $\SM$ to the supertime $\SM_0$ is given in coordinates by $\pi: (t,\theta,\bar\theta,x^i)\mapsto (t,\theta,\bar\theta)$.  We will sometimes refer to the coordinates $(t,\theta,\bar\theta)$ on supertime collectively as $\tau^M$, with the coordinate index $M\in\{t,\theta,\bar\theta\}$.

Note that the dimensions $\theta$ and $\bar\theta$ are two real Grassmannian coordinates, and they supersymmetrize only the time dimension; the bosonic spatial coordinates $x^i$ parametrize the leaves of the foliation, and can often be viewed as spectators from the perspective of the supersymmetrized time.  In what follows, we will use interchangeably $\p_t$ and $\dot{\ }$ to denote the time derivative $\p/\p t$.

The theories we will be interested in will exhibit $\CN=2$ supersymmetry,%
\footnote{In our conventions, $\CN$ counts the number of individual real supercharges.}
 with supercharges realized on $\SM$ as differential operators
\be
Q=\frac{\p}{\p\theta},\qquad \bar Q=\frac{\p}{\p\bar\theta}+\theta\p_t,
\ee
and satisfying the superalgebra
\be
\{Q,\bar Q\}=\p_t,\qquad Q^2=\bar Q^2=0.
\label{eealgr}
\ee
In this $\CN=2$ superalgebra, we intend to identify $Q$ to be our BRST charge.  Thus, physical states and operators in our topological gravity theory will be determined from the cohomology of $Q$.  However, for the time being we suspend this underlying BRST interpretation, and simply construct our theory as a supersymmetric theory with the rigid $\CN=2$ superalgebra (\ref{eealgr}).  

The superderivatives that anticommute appropriately with the supercharges are:
\be
\RD =\frac{\p}{\p\theta}-\bar\theta\p_t,\qquad \barRD =\frac{\p}{\p\bar\theta};
\ee
they satisfy
\be
\{\RD,\barRD\}=-\p_t,\qquad \RD^2=\barRD^2=0,
\ee
and
\be
\{\RD,Q\}=\{\RD,\bar Q\}=\{\barRD,Q \}=\{\barRD,\bar Q \}=0,
\ee
and of course $\RD$ and $\barRD$ both commute with the spatial derivative $\p_i\equiv\p/\p x^i$.  

\subsection{The action}
\label{secactprim}

Our primitive theory is a topological theory of the component fields contained in the spatial metric superfield $G_{ij}$.  It will have no gauge symmetries, and it will respect the $\CN=2$ supersymmetry algebra described above.  In addition, we will require that the theory be invariant under time-independent spatial diffeomorphisms of the spatial slices $\Sigma$.  Since this symmetry does not depend on time, it is better not to interpret it as a gauge symmetry, despite its dependence on the location along $\Sigma$.  In the present setting, $\mathrm{Diff}(\Sigma)$ essentially represents an infinite-dimensional global symmetry.

Under these symmetry assumptions, we now write the superspace action as a sum of two terms,
\be
S=\frac{1}{\kappa^2} \left(S_K-S_\CW\right).
\label{eeoriga}
\ee
The kinetic term $S_K$ is a sum of all the invariants that contain at least one supertime derivative, while the potential term $S_\CW$ contains all the invariants with only spatial derivatives but no supertime derivatives.  In the component form, this decomposition will translate into $S_K$ containing at least one time derivative, and $S_\CW$ including all the terms without time derivatives. Extending the customary physics terminology to this case, we will refer to $S_\CW$ as the ``superpotential.'' Both $S_K$ and $S_\CW$ are integrals of a local Lagrangian density over all of superspace,%
\footnote{As usual, we define the measure $d^2\theta$ and the Berezin integral over the anticommuning coordinates by linearity together with $\int d^2\theta\,\theta\bar\theta=1$ and $\int d^2\theta\,\theta=\int d^2\theta\,\bar\theta=\int d^2\theta\,1=0$.}
\be
S_K=\int dt\,d^Dx\,d^2\theta\,\CL_K,\qquad S_\CW=\int dt\,d^Dx\,d^2\theta\,\CL_\CW.
\label{eeactgen}
\ee
We will require that they preserve the ghost number symmetry, $\gh(S_K)=\gh(S_\CW)=0.$  Note that for future convenience, we have factored out one overall coupling constant, $\kappa^2$, in front of the entire action.  

The terms that can appear in $S_K$ and $S_\CW$ can be usefully organized by their increasing classical scaling dimensions.  Until or unless we commit to a particular value of the dynamical scaling exponent $z$, time and space scaling is unrelated (as we reviewed briefly at the beginning of Section~\ref{ssprim}), and we assign classical scaling dimensions to the ingredients appearing in the action as follows: $[\p_i]=L^{-1}, [\p_t]=T^{-1}, [G_{ij}]=0$.  The superalgebra implies that $[\RD ]+[\barRD ]=T^{-1}$.%
\footnote{Sometimes, in various dynamical regimes, it is convenient to choose a specific value of the dynamical exponent $z$, which relates the scaling of time and space, so that $T$ scales as $T\sim L^z$.  In that case it is then conventional to assign the classical ``scaling dimension'' $\Delta$ to any object $\CO$ if $\CO$ scales as $T^{-\Delta}$, \ie , to measure the scaling dimension in the units of energy.  Also, since in this paper we are focusing on the basic set-up of the path integral representation of the theory, and do not calculate any quantum corrections to classical scaling dimensions, all our scaling dimensions will be classical.  We will follow these conventions throughout.}
The terms in $S_K$ of the lowest scaling dimension (\ie , with the lowest number of derivatives) will be of dimension $T^{-1}L^0$.  The first obvious candidate would be $\int\sqrt{G}G^{ij}\dot G_{ij}$, but that term is a total derivative, $\int\p_t(2\sqrt{G})$, and hence gives no local dynamics.  A nontrivial leading-order kinetic term of this dimension can indeed be constructed; it contains two superderivatives,
\be
\label{eeactsk}
S_K=\int dt\,d^2\theta\,d^D x\,\sqrt{G}\left\{\left(\lambda_\perp G^{ik}G^{j\ell}-\lambda G^{ij}G^{k\ell}\right)\barRD G_{ij}\,\RD G_{k\ell}+\ldots\ \right\}
\ee
(The ``$\ldots$'' stand as a reminder that there may be terms of higher scaling dimension that one may wish to include.)  This kinetic term depends on two coupling constants $\lambda$ and $\lambda_\perp$, which we take to be of scaling dimension zero: $L^0T^0$.  This in turn implies that the scaling dimension of $\kappa^2$ is $[\kappa^2]=L^D T^{-1}$.  Clearly, $\lambda_\perp$ is redundant, and one usually sets  $\lambda_\perp=1$.  We will do so from now on, but we wish to point out that the implicit assumption leading to this step is that $\lambda_\perp$ is positive, while in some circumstances these types of theories can also be studied in the regime where $\lambda_\perp\leq 0$.  

The superpotential terms can be similarly organized by the number of increasing spatial derivatives.  Focusing on the terms with up to two derivatives, we find two terms respecting all our global symmetries: the Ricci scalar of $G_{ij}$ and the cosmological constant term,
\be
\label{eeactsw}
S_\CW=\int dt\,d^2\theta\,d^D x\,\sqrt{G}\left\{\ldots +\alpha_R R^{(G)}+\alpha_\Lambda\right\}.
\ee
We will always refer to the various couplings in the superpotential as $\alpha,$ with an appropriate subscript indicating the term each particular coupling is associated with.  Thus, here $\alpha_R$ is the coupling associated with the spatial Einstein-Hilbert term in superspace, and $\alpha_\Lambda$ is the superspace cosmological constant.  We organized the terms in the order of their increasing scaling dimension from the right to the left, with the ``$\ldots$'' on the left standing for all terms with more than two derivatives.  We of course assume the perspective and logic of effective quantum field theory here, implying that all terms consistent with the underlying symmetries are in principle present.  In some cases, only a finite number of terms up to a certain ``critical'' dimension is sufficient to make the theory perturbatively renormalizable, or perhaps even short-distance complete, without the need for higher-derivative terms.  The analysis of possible short-distance completeness of the topological gravity theories presented in this paper is a fascinating open question for future research.%
\footnote{The naive scaling properties of free-field fixed points suggest that in $3+1$ dimensions, power-counting renormalizability requires $z=3$, implying in turn that all terms up to three derivatives would need to be included in the superpotential.  This would include the gravitational Chern-Simons 3-form built out of the Levi-Civit\`a connection of $G_{ij}$, and would lead to a generalization of the Ricci flow involving the Cotton tensor \cite{lif,cflow,lmcflow}.  In $2+1$ dimensions, only terms up to two derivatives are sufficent for power-counting renormalizability.}

Now we are ready to see the relation between our primitive supersymmetric theory and the Ricci flow equations.  We perform the $d^2\theta$ integral in $S_K$ and $S_\CW$ to obtain the action in component form.  In components, the action (\ref{eeoriga}) with $S_K$ given in (\ref{eeactsk}) and with a general superpotential term $S_\CW$ given by (\ref{eeactgen}) takes the following form,
\be
S=\frac{1}{\kappa^2}\int dt\,d^D x\,\left\{ \sqrt{g}(g^{ik}g^{j\ell}-\lambda g^{ij}g^{k\ell})B_{ij}\left(\dot g_{k\ell}-B_{k\ell}\right)-B_{ij}\frac{\delta\CF}{\delta g_{ij}}+ \mathrm{fermions}\right\}.
\label{eeprimcomp}
\ee
Here we defined $\CF$ to be the (bosonic) spacetime integral of the lowest component $\CL_\CF$ of the $\CL_\CW$ superfield in the $\theta,\bar\theta$ expansion: 
\bea
\CL_\CW&=&\CL_\CF+\textrm{higher orders in}\ \theta,\bar\theta,\\
\CF&=&\int dt\,d^Dx\,\CL_\CF.
\eea
The auxiliary field $B_{ij}$ can be integrated out, and the bosonic part of the action then becomes
\bea
S_{\mathrm{bose}}&=&\frac{1}{4\kappa^2}\int dt\,d^D x\,\sqrt{g}(g^{ik}g^{j\ell}-\lambda g^{ij}g^{k\ell})\left[\dot g_{ij}-\frac{1}{\sqrt{g}}(g_{im}g_{jn}-\tilde\lambda g_{ij}g_{mn})\frac{\delta\CF}{\delta g_{mn}}\right]\\
&&\qquad\qquad\qquad{}\times\left[\dot g_{k\ell}-\frac{1}{\sqrt{g}}(g_{kr}g_{\ell s}-\tilde\lambda g_{k\ell}g_{rs})\frac{\delta\CF}{\delta g_{rs}}\right],
\eea
where $\tilde\lambda$ is given, as usual in Lifshitz gravity \cite{mqc}, by
\be
\tilde\lambda=\frac{\lambda}{D\lambda-1}.
\ee
Clearly, for values of $\lambda\leq 1/D$, this action is bounded from below by zero, and this bound is saturated when the metric satisfies the appropriate flow equation, of first order in time derivatives.

The fermionic component contributions to (\ref{eeprimcomp}) are straightforward to determine, but they look a little cumbersome and we suppress them for the ease of the presentation, as is often done in supergravity theories.  Perhaps the most important thing to remember about the fermions is that they also have a non-degenerate kinetic term,
\be
\chi_{ij}(g^{ik}g^{j\ell}-\lambda g^{ij}g^{k\ell})\dot\psi_{k\ell}+\ldots,
\ee
and therefore our entire theory can be treated in a perturbative expansion using standard Feynman diagram techniques.  

The quantum theory of our primitive topological gravity is formally defined via the path integral as a sum over all appropriate histories,
\be
\CZ=\int\SD\mu[G_{ij}]\,\exp\left\{-\frac{1}{\hbar\kappa^2}(S_K-S_\CW)\right\}.
\label{pathintp}
\ee
Here $\SD\mu[G_{ij}]$ is the $\CN=2$ supersymmetric measure on the space of the metric fields.  

A few comments about some salient features of this path integral seem in order.  Many of them will be relevant also to the more sophisticated cousins of the primitive theory, which we will develop below.

\begin{itemize}
\item
  In order to become well-defined even by the physics standard of rigor, this path integral requires that appropriate boundary conditions be specified at the boundaries of spacetime, even in the absence of singularities.  What is the correct question to ask must be guided by physics principles:  We must first decide what is the appropriate set of probability amplitudes and observables that are meaningful in this context of time-dependent quantum gravity and cosmology.  We might be interested in choosing the initial surface $\Sigma$ and calculating the Hartle-Hawking-type wavefunction of the Universe.  Or perhaps one might wish to evaluate the transition amplitudes between physical states at an initial and finite time.  Besides calculating the partition function $\CZ$ or transition amplitudes with such boundary conditions, one can define correlation functions of BRST-invariant local operators, or of observables associated with extended submanifolds in spacetime.  This question of observables is beyond the scope of the present paper, but represents an intriguing opportunity to find a new window into quantum gravity and quantum cosmology far from equilibrium, at least in the topological setting.  
\item
  Standard arguments of topological quantum field theory apply \cite{ewcoho,ewtym}, at least formally: The overall coupling $\kappa^2$ plays the role of $\hbar$.  The semiclassical approximation at small $\kappa$ is ``exact'' at one loop, and the path integral localizes to the space of solutions of the localization equation, which in our case is a Ricci-type flow equation for $g_{ij}$.  A similar argument implies that the physical observables (such as the partition function) are independent of the small changes in the coupling constants; here ``small'' means roughly those changes which do not lead to degeneracies in the action.
\item
  Note that our theory is formally defined in ``imaginary time''.  One might also be interested instead in the ``real-time'' path integral, which would have the integrand $\exp (iS)$ instead of the $\exp (-S)$ appearing in (\ref{pathintp}).  This possibility is already interesting for the primitive theory, but will become even more relevant for the more sophisticated versions of topological quantum gravity constructed below, which have some form of spacetime diffeomorphism invariance.  We will further comment on this possibility of continuing to real time in Section~\ref{ssreal}.
\item
  Already this simplest ``primitive'' theory depends on several coupling constants: $\lambda,\alpha_R$ and $\alpha_\Lambda$ (and perhaps more, if we choose to add higher-derivative terms), and the classical localization equations thus represent a multi-parameter generalization of the standard Ricci flow.  It will be important to subject this ``landscape'' of topological gravity theories to a closer study, to see what limits on the values of the coupling constants naturally emerge from requiring that the formal path integral satisfy various physical consistency conditions.  In particular, not for all values of the couplings will the solutions of the localization equations be as well-behaved as those of Hamilton's Ricci flow, putting bounds on the range of the couplings.  Of course, this broader family of generalized flow equations has been much less studied in the mathematical literature, and much less is known exactly.
\end{itemize}

\subsection{Localization and Hamilton's Ricci flow}
\label{sshamloc}

To see that Hamilton's original Ricci flow indeed appears in the landscape of our theory, let us take a closer look at the localization equation, obtained from (\ref{eeprimcomp}):
\be
\frac{\p g_{ij}}{\p t}=\frac{1}{\sqrt{g}}(g_{ik}g_{j\ell}-\tilde\lambda g_{ij}g_{k\ell})\frac{\delta\CF}{\delta g_{k\ell}},
\ee
With the specific form of the superpotential given in (\ref{eeactsw}), this becomes
\be
\frac{\p g_{ij}}{\p t}=-\alpha_RR_{ij}+\frac{\alpha_R}{2}\left[1-\tilde\lambda(D-2)\right]g_{ij}R+\frac{\alpha_\Lambda}{2}g_{ij}.
\label{eelocprim}
\ee
We observe that setting
\be
\alpha_R=2,\qquad\alpha_\Lambda=0,\qquad\lambda=\frac{1}{2}
\label{eecoupsh}
\ee
in the action of the primitive theory reduces the localization equation (\ref{eelocprim}) to
\be
\frac{\p g_{ij}}{\p t}=-2R_{ij}.
\ee
Thus, for the values of the couplings given in (\ref{eecoupsh}), the original Ricci flow equation (\ref{eehamrf}) of Hamilton's appears as the localization equation in our theory of topological quantum gravity, despite the fact that it is not a gradient flow equation.

Interestingly, the value $\lambda=1/2$ that leads to Hamilton's Ricci flow is \textit{not} in the range of $\lambda$ in which the action is positive definite.  The proper treatment of the path integral would require a rather subtle analytic continuation.  If we wanted to make sense of this continuation, we would be facing a situation very analogous to relativistic Euclidean quantum gravity \cite{eucl}, in which the Euclidean action is also not bounded from below, due to the contributions from the spacetime scale factor of the metric.  In the nonrelativistic context relevant here, the culprit is the scale factor of the spatial metric $g_{ij}$. 

\section{The gauge theory:  Gauging spatial diffeomorphisms}
\label{ssdiffs}

In the next step, we wish to incorporate some of the gauge symmetries expected of quantum gravity into our topological theory.  We begin by gauging the time-independent symmetries of spatial diffeomorphisms $\mathrm{Diff}(\Sigma)$ exhibited by the primitive theory.

The basic field of the primitive theory was the spatial metric $g_{ij}(t,x^k)$.
Under an infinitesimal time-dependent spatial diffeomorphism $\xi^i(t,x^j)$, the metric tensor would transform as
\be
\label{metdifb}
\delta g_{ij}=\xi^k\p_kg_{ij}+g_{kj}\p_i\xi^k+g_{ik}\p_j\xi^k.
\ee
Note that in this relation, the time coordinate plays the role of a spectator:  not only the time-independent but also the time-dependent spatial diffeomophisms act via (\ref{metdifb}) leaf-by-leaf, at each fixed $t$, as ordinary spatial diffeomorphisms.

In the primitive theory, we supersymmetrized the spatial metric by promoting $g_{ij}(t,x^k)$ to an unconstrained $\CN=2$ superfield $G_{ij}(t,\theta,\bar\theta,x^k)$, whose component expansion we recall here,
\be
G_{ij}=g_{ij}+\theta\psi_{ij}+\bar\theta\chi_{ij}+\theta\bar\theta B_{ij}.
\ee
In order to gauge the spatial diffeomorphisms consistently with the $\CN=2$ supersymmetry, we follow the strategy familiar from supersymmetric Yang-Mills theories in superspace:  We promote the diffeomorphism generator $\xi^i$ into a superfield,
\be
\Xi^i=\xi^i+\theta\zeta^i+\bar\theta\eta^i+\theta\bar\theta\alpha^i.
\label{eexisym}
\ee
(Later on, we will impose various chirality constraints on $\Xi^i$, but for now we will treat it as unconstrained.)  Under the spatial superdiffeomorphisms, the metric superfield $G_{ij}$ transforms in a straightforward generalization of (\ref{metdifb}), as
\be
\label{metdif}
\delta G_{ij}=\Xi^k\p_kG_{ij}+G_{kj}\p_i\Xi^k+G_{ik}\p_j\Xi^k.
\ee
Note that in this transformation rule, both $t$ and $\theta,\bar\theta$ again play the role of spectators, and (\ref{metdif}) acts at each fixed value of the specator supercoordinates (defining an individual leaf of the foliation) as a spatial superdiffeomorphism of the metric superfield along the leaf.

\subsection{ABCs of supersymmetrizations of the Diff($\Sigma$) symmetry}

We begin with the primitive theory, and $G_{ij}$ as the only superfield.  In order to promote the transformations of the spatial $\diff(\Sigma)$ symmetries into a gauge symmetry, one must introduce the appropriate gauge fields which allow us to covariantize the time derivatives of the spatial metric.

Let us first recall how this works in bosonic gravity.  The role of such gauge fields is played by the famous ``shift vector'' $n^i$ (in the terminology of the ADM formalism), which transforms as
\be
\label{ndifb}
\delta n^i=\dot\xi^i+\xi^k\p_kn^i-n^k\p_k\xi^i.
\ee
The interpretation of the three terms in (\ref{ndifb}) is very clear:  The first term, viewed for each fixed value of the spatial index $i$ is exactly the transformation of an Abelian gauge field under a time-dependent gauge transformation with parameter $\xi^i$, with one such Abelian symmetry for each spatial dimension.  And the second plus third term are a nonlinear correction to this leading gauge transformation, which ensure that $\xi^i$ are not independent Abelian symmetries but represent the nonlinear transformations of spatial diffeomorphisms.  These two terms make sure that $n^i$ transform correctly as components of a spatial one-vector under time-independent spatial diffeomorphisms.  

Using the shift vector $n^i$, one can now covariantize $\dot g_{ij}$ to
\be
\nabla_t g_{ij}
\equiv\dot g_{ij}-n^k\p_kg_{ij}-g_{kj}\p_in^k-g_{ik}\p_jn^k,
\ee
and show that this covariantization transforms correctly, as a spatial two-tensor, under time-dependent spatial diffeomorphisms $\xi^i(t,x^k)$:
\be
\delta (\nabla_t g_{ij})=\xi^k\p_k(\nabla_t g_{ij})+(\nabla_t g_{kj})\p_i\xi^k+(\nabla_t g_{ik})\p_j\xi^k.
\ee

The supersymmetrization of the covariant time derivative is straightforward:  We promote the shift vector $n^i$ into a superfield 
\be
N^i=n^i+\theta\psi^i+\bar\theta\chi^i+\theta\bar\theta B^i,
\label{eesupersh}
\ee
and postulate that $N^i$ transform under superdiffeomorphisms $\Xi^i$ as
\be
\label{ndif}
\delta N^i=\dot\Xi^i+\Xi^k\p_kN^i-N^k\p_k\Xi^i.
\ee
For now, we treat $N^i$ as an unconstrained superfield, but will see below that it might be consistent with various chirality constraints.  We extend the definition of the covarantized time derivative $\nabla_t$ to the superfield $G_{ij}$, 
\be
\nabla_tG_{ij}\equiv \dot G_{ij}-N^k\p_kG_{ij}-G_{kj}\p_iN^k-G_{ik}\p_jN^k,
\ee
and observe that the superfield $\nabla_tG_{ij}$ transforms under $\Xi^i$ as a spatial two-tensor,
\be
\delta (\nabla_tG_{ij})=\xi^k\p_k(\nabla_tG_{ij})+(\nabla_tG_{kj})\p_i\xi^k+(\nabla_tG_{ik})\p_j\xi^k.
\ee
Indeed, this is a simple consequence of (\ref{ndifb}) together with the fact that $\theta,\bar\theta$ play the role of spectators in our construction of the covariant time derivative.  

Having covariantized the time derivative of $G_{ij}$, we must now covariantize the superderivatives $\RD G_{ij}$ and $\barRD G_{ij}$.  We first introduce gauge superfields $S^i$ and $\bar S^i$, of the opposite statistics to $N^i$, and such that they transform under the gauge supertransformations $\Xi^i$ as
\bea
\label{sdif}
\delta S^i&=&\RD\Xi^i+\Xi^k\p_kS^i-S^k\p_k\Xi^i,\\
\label{sbardif}
\delta \bar S^i&=&\barRD\Xi^i+\Xi^k\p_k\bar S^i-\bar S^k\p_k\Xi^i.
\eea
With such superconnections, we now define the covariantized superderivatives of $G_{ij}$,
\bea
\CD G_{ij}&\equiv& \RD G_{ij}-S^k\p_kG_{ij}-G_{kj}\p_iS^k-G_{ik}\p_jS^k,\\ 
\bar\CD G_{ij}&\equiv& \barRD G_{ij}-\bar S^k\p_kG_{ij}-G_{kj}\p_i\bar S^k-G_{ik}\p_j\bar S^k,
\eea
and see that they transform correctly under $\Xi^i(t,\theta,\bar\theta,x^k)$, as spatial two-tensors.

\subsubsection{Type C: The chiral theory}
\label{sectypec}

Before studying in more detail this general case, we first observe that one can consistently restrict $\Xi^i$ to be chiral superfields,
\be
\barRD\Xi^i=0.
\ee
This will define what we will refer to as ``Type C theory'' (here ``C'' naturally stands for ``chiral'').  In Type C theory, the ordinary superderivative $\barRD$ is already covariant, and no $\bar S^i$ superconnection is needed.  Only $S^i$ must be introduced, to covariantize $\RD$ into $\CD$.  Still, having both $N^i$ and $S^i$ without any relation between them would lead to too many gauge field components (for example, both $N^i$ and $S^i$ contain a bosonic component that transforms as the bosonic shift vector $n^i$).  In order to find a suitable constraint that relates them, note first that $-\barRD S^i$ transforms as $N^i$.  This leads us to expect that in Type C theory, 
\be
\label{constns}
N^i=-\barRD S^i.
\ee
In fact, this constraint has a very clear geometric origin, closely reminiscent of similar constraints in supersymmetric Yang-Mills gauge theories:  It simply states that the action of the anticommutator of the covariantized superderivatives $\{\CD,\barRD\}$ on $G_{ij}$ (or, indeed, on any symmetric 2-tensor $T_{ij}$) in Type C theory reproduces the action of $-\nabla_t$ on $G_{ij}$ (or $T_{ij}$).  Here we must be careful of the order of terms when evaluating $\CD$ of an odd two-tensor; the correct formula that works regardless of the statistics of $T_{ij}$ is
\be
\CD T_{ij}\equiv \RD T_{ij}-\p_iS^k T_{kj}-\p_jS^k T_{ik}-S^k\p_k T_{ij}.
\ee
Note that since $N^i$ in Type C theory is $\barRD$ of something, it is automatically chiral:
\be
\barRD N^i=0.
\ee
This is pleasing, since such a chiral $N^i$ (for each fixed $i$) contains one real bosonic component $n^i$ and one real fermionic component, which matches the number of independent component gauge transformation contained in a chiral $\Xi^i$.  We can thus plan on eliminating the fermionic component of $N^i$ by going to the analog of Wess-Zumino gauge \cite{wessz}.

We would similarly expect that $S^i$ should have only two independent components in the $\theta,\bar\theta$ expansion.  However, we clearly cannot impose the antichirality condition and simply set $\RD S^i$ to be zero:  This would be inconsistent with the fact that under a chiral $\Xi^i$, the transformation $\delta S^i$ is {\it not\/} antichiral (even though it would be so at the linearized level).  So, either $S^i$ is unconstrained, and therefore contains four independent components two of which would have to be gauge invariant (which would be unpleasant), or there is another constraint that can be consistently imposed on $S^i$.  The correct constraint turns out to be nonlinear,
\be
\RD S^i=S^k\p_kS^i,
\ee
and it represents a covariant version of the antichirality condition.  We will return to its precise geometric interpretation in Section~\ref{secgeofconn}.

\subsubsection{Type A: The antichiral theory}

Instead of postulating that the spacetime superdiffeomorphisms $\Xi^i$ are chiral as in Type C theory, we could start with the antichirality condition,
\be
\RD\Xi^i=0.
\ee
The entire construction will go through in the same way as in Type C theory, with all chiralities and antichiralities reversed at all the relevant steps.  We will refer to this construction as ``Type A theory'' (with ``A'' standing for ``antichiral'').  Since the theory enjoys $\CN=2$ supersymmetry, Type A theory might naively seem like another construction of the same Type C theory in disguise, up to a simple change of coordinates.  However, recall that when we introduced our supercharges, we selected once and for all $Q$ (and not $\bar Q$, or any other linear superposition of them) to be our BRST charge of topological symmetry.  This selection lifts the $\CN=2$ democracy between the two chiralities, and makes Type A theory {\it a priori\/} distinct from Type C.

In more detail, in Type A theory we covariantize the time derivative using gauge superfield $N^i$, which transforms according to (\ref{ndif}), now with an antichiral $\Xi^i$.  And we covariantize the superderivative $\barRD$ to $\bar\CD$ by introducing the odd gauge superfield $\bar S^i$.  The other superderivative $\RD$ is already covariant, and no $S^i$ superfield is introduced or needed.  The relation between $N^i$ and $\bar S^i$ in Type A theory is
\be
N^i=-\RD\bar S^i,
\ee
which makes $N^i$ automatically antichiral.  This relation is again an expression of a covariant constraint, which ensures that 
\be
\{\RD,\bar\CD\} T_{ij}=-\nabla_t T_{ij},
\ee
on any symmetric 2-tensor $T_{ij}$.

Note that $\bar S^i$, if further unconstrained, would have two gauge-invariant components, for which we have no use.  A constraint should again be imposed to eliminate them, but it cannot be simply the chirality condition on $\bar S^i$, which is inconsistent with the transformations of $\bar S^i$ under antichiral superdiffeomorphisms.  The correct constraint takes the form of a nonlinear improvement of the naive antichirality constraint,
\be
\barRD\bar S^i=\bar S^k\p_k\bar S^i.
\ee

\subsubsection{Type B: The balanced theory}
\label{secb}

While Theories C and A appear to be the minimal theories with $\CN=2$ supersymmetry and gauge superdiffeomorphism symmetry, they each break the symmetry between ghosts and antighosts, due to the (anti)chirality condition on the superdiffeomorphism parameters $\Xi^i$.  Now we will construct a theory with $\Xi^i$ fully unconstrained, which will restore the ghost-antighost symmetry.  In the literature, topological theories with such a ghost-antighost symmetry are sometimes referred to as ``balanced'' \cite{dmbala}.  We will adopt this terminology for our case here as well, and will call this theory ``Type B'' (with ``B'' naturally standing for ``balanced'').  

In order to allow for unconstrained $\Xi^i$ supergauge transformations, we must covariantize the time derivative and both superderivatives $\RD$, $\barRD$, by introducing gauge superfields $N^i$, $S^i$ and $\bar S^i$, which transform according to (\ref{ndif}), (\ref{sdif}) and (\ref{sbardif}).  They correctly covariantize all our derivatives, but carry way too many independent components and must therefore be subjected to a series of natural constraints.  First of all, $N^i$ can be algebraically expressed in terms of $S^j$, $\bar S^j$ and their various derivatives, by imposing
\be
\label{ttbarconst}
\left\{\CD,\bar\CD\right\} T_{ij}=-\nabla_t T_{ij}
\ee
on symmetric two-tensors.  This condition gives
\be
\label{nssbar}
N^i=-\barRD S^i-\RD\bar S^i+S^k\p_k\bar S^i+\bar S^k\p_kS^i.
\ee
Note several interesting facts about this formula:  First of all, in the absence of $\bar S^k$ (or $S^k$), it reduces to the expressions for $N^i$ in Type C (or Type A) theory, respectively.  Secondly, in the Type B theory, the expression for $N^i$ also contains important {\it nonlinear cross-terms\/} between $S^i$ and $\bar S^k$, which had no analog in Type C and Type A theories.

Our relation (\ref{nssbar}) uniquely expresses $N^i$ in terms of $S^i$ and $\bar S^i$.  Thus, we expect that Wess-Zumino gauge exists, in which we keep only the leading component $n^i$ and the bosonic diffeomorphisms $\xi^i$ as symmetries, using the remaining three components of $\Xi^i$ (for each $i$) to eliminate the remaining three components of $N^i$.  However, this still leaves us with too many components of the {\it a priori\/} unrelated $S^i$ and $\bar S^i$, a problem which we already noticed in Type C and Type A theories.  We therefore return to the geometric interpretation of all our constraints, in the ``umbrella'' case of Type B theory.

\subsection{Geometric interpretation I: Superconnections, constraints and flatness}
\label{secgeofconn}

To find suitable constraints that should be imposed on $S^i$ and $\bar S^i$, we can calculate the appropriate graded commutators of our covariant derivatives, and define ``supercovariant field strengths'' $W_{MN}^i$, $M,N\in\{t,\theta,\bar\theta\}$, in a way reminiscent of more traditional supersymmetric gauge theories (such as super Yang-Mills), as obstructions against the closure of the algebra of derivatives isomorphic to the algebra of $\RD,\barRD$ and $\p_t$.  Evaluating the graded commutators of $\CD,\bar \CD$ and $\nabla_t$ on our spatial metric superfield $G_{ij}$ gives:
\bea
\{\CD,\CD\}G_{ij}&=& -\p_iW_{\theta\theta}^kG_{kj}-\p_jW_{\theta\theta}^kG_{ik}-W_{\theta\theta}^k\p_kG_{ij},\\
\{\bar\CD,\bar\CD\}G_{ij}&=& -\p_iW_{\bar\theta\bar\theta}^kG_{kj}-\p_jW_{\bar\theta\bar\theta}^kG_{ik}-W_{\bar\theta\bar\theta}^k\p_kG_{ij},\\
\{\CD,\bar\CD\}G_{ij}&=& -\nabla_tG_{ij}- \p_iW_{\theta\bar\theta}^kG_{kj}-\p_jW_{\theta\bar\theta}^kG_{ik}-W_{\theta\bar\theta}^k\p_kG_{ij},\label{eethird}\\
\left[\nabla_t,{\CD}\right] G_{ij}&=&-\p_iW_{t\theta}^kG_{kj}- \p_jW_{t\theta}^kG_{ik}-W_{t\theta}^k\p_kG_{ij},\\
\left[\nabla_t,{\bar\CD}\right] G_{ij}&=&-\p_iW_{t\bar\theta}^kG_{kj}- \p_jW_{t\bar\theta}^kG_{ik}-W_{t\bar\theta}^k\p_kG_{ij},
\eea
where
\bea
W_{\theta\theta}^i&=&2(\RD S^i-S^k\p_kS^i),\label{eestrone}\\
W_{\bar\theta\bar\theta}^i&=&2(\barRD\bar S^i-\bar S^k\p_k\bar S^i),\\
W_{\theta\bar\theta}^i&=&N^i+\barRD S^i+\RD\bar S^i-\bar S^k\p_kS^i-S^k\p_k\bar S^i,\\
W_{t\theta}^i&=&\dot S^i-N^k\p_kS^i-\RD N^i+S^k\p_kN^i,\\
W_{t\bar\theta}^i&=&\dot{\bar S}^i-N^k\p_k\bar S^i-\barRD N^i+\bar S^k\p_kN^i.\label{eestrfive}
\eea
Note that our constraint (\ref{nssbar}) is simply indicating the vanishing of the field strength
\be
\label{ttbarw}
W_{\theta\bar\theta}^i=0.
\ee
This suggests that we impose the rest of the ``flatness conditions'' on supertime,
\be
\label{ttw}
W_{\theta\theta}^i=0,\qquad W_{\bar\theta\bar\theta}^i=0,
\ee
which requires
\be
\label{sconst}
\RD S^i=S^k\p_kS^i,\qquad\barRD \bar S^i=\bar S^k\p_k\bar S^i.
\ee
In fact, our set of constraints (\ref{nssbar}) and (\ref{sconst}) is the minimal set that implies the vanishing of {\it all\/} the field strengths $W_{MN}^i$ by Bianchi identities.  

In our construction of Type C and A theories, we pointed out that $S^i$ and $\bar S^i$ should each satisfy a chirality-like constraint, but showed the inconsistency of constraining $S^i$ and $\bar S^i$ by the naive linear (anti)chirality conditions.  The two constraints (\ref{sconst}) are the required consistent nonlinear extensions of the naive (anti)chirality constraints.  Note that these two conditions remain nonlinear in $S^i$ (or $\bar S^i$) when reduced to Type C (or Type A) theory.  

Any potential worry that the nonlinear nature of the constraints (\ref{sconst}) could lead to over-constraining is eliminated by finding the explicit solutions of the constraints in components.  Constraints (\ref{sconst}) are solved by
\bea
S^i&=&\sigma^i+\theta\sigma^k\p_k\sigma^i+\bar\theta Y^i+\theta\bar\theta(\dot\sigma^i+Y^k\p_k\sigma^i-\sigma^k\p_kY^i),\\
\bar S^i&=&\bar\sigma^i+\theta X^i+\bar\theta\bar\sigma^k\p_k\bar\sigma^i+\theta\bar\theta(\bar\sigma^k\p_k X^i-X^k\p_k\bar\sigma^i).
\eea
Then (\ref{nssbar}) is solved by setting
\be
Y^i=-n^i-X^i+\sigma^k\p_k\bar\sigma^i+\bar\sigma^k\p_k\sigma^i,
\ee
and expressing the three remaining components in the $N^i$ superfield (\ref{eesupersh}) as follows,
\bea
\psi^i&=&\dot\sigma^i+\sigma^k\p_k n^i-\p_k\sigma^i n^k,\\
\chi^i&=&\dot{\bar\sigma}^i+\bar\sigma^k\p_k n^i-\p_k\bar\sigma^i n^k,\\
B^i&=&-\dot X^i+n^k\p_k X^i-X^k\p_kn^i+\dot\sigma^k\p_k\bar\sigma^i+\sigma^j\p_jn^k\p_k\bar\sigma^i-n^j\p_j\sigma^k\p_k\bar\sigma^i\nonumber\\
&&\ {}+\bar\sigma^k\p_k\dot\sigma^i+\bar\sigma^k\p_k\sigma^j\p_jn^i+\bar\sigma^k\sigma^j\p_k\p_jn^i-\bar\sigma^k\p_kn^j\p_j\sigma^i-\bar\sigma^kn^j\p_k\p_j\sigma^i.
\eea
All the constraints have been solved, and the shift superfields are all expressed in terms of the bosonic component fields $n^i,X^i$, and the fermionic components $\sigma^i$ and $\bar\sigma^i$.

This concludes our construction of the self-consistent covariantization under unconstrained diffeomorphisms $\Xi^i$. 
To summarize, imposing the flatness conditions (\ref{ttbarw}) and (\ref{ttw}) leads to the set of constraints (\ref{nssbar}) and (\ref{sconst}) on $N^i$, $S^i$ and $\bar S^i$.  This reduces the number of independent components in the gauge superfields so that we can eliminate all the components besides $n^i$ by an analog of the Wess-Zumino gauge.  

\subsection{Geometric interpretation II: Supersymmetric Diff($\Sigma$) Yang-Mills theory}

The set of constraints that we just identified in terms of the superfield strengths has another intriguing geometric interpretation, which sheds some additional light on our covariantization construction, and which can also be of independent interest.

In this new interpretation, we take a different perspective on the structure of spacetime:  We view the theory as a supersymmetric gauge theory on supertime $\SM_0$, \ie , a theory in $(1|2)$ dimensions.  The spatial slices $\Sigma$ will be interpreted as an internal space, not as dimensions of spacetime.  Thus, a field such as $N^i(t,\theta,\bar\theta,x^j)$ is interpreted as a field on $(t,\theta,\bar\theta)$, with $(i,x^i)$ interpreted as a continuous internal (multi)index.

Recall that for any (typically compact, and finite-dimensional) internal Yang-Mills symmetry group $\CG$, there are standard rules for constructing the corresponding supersymmetric Yang-Mills theory in a superspace of dimension $(d|d')$: One postulates the existence of superconnections $\Gamma^\alpha_M$ on superspace, where the index $M$ goes over all $d+d'$ values, and $\alpha$ indicates the adjoint representation of the bosonic gauge group $\CG$.  The derivatives on superspace have thus been covariantized to $D_M$.  Next one defines the supersymmetric field strengths $W_{MN}^\alpha$ via
\be
[D_M,D_N\}=T_{MN}{}^PD_P+W_{MN},\qquad W_{MN}^\alpha T_\alpha,
\label{eedefgw}
\ee
with $T_{MN}{}^P$ the torsion of the flat superspace, and $T_\alpha$ the generators of $\CG$.  Finally, one imposes a set of constraints sometimes referred to as ``conventional'' , \cite{west,wb}:
\be
W^\alpha_{MN}=0,\qquad\mathrm{whenever\ both\ indices}\ M,N\ \mathrm{are\ odd}.
\label{eeconve}
\ee
The rest of the constraints is implied by the Bianchi identities.

With the full list of constraints identified, one can then construct various candidate Lagrangians in superspace, typically by invoking an invariant metric $g_{\alpha\beta}$ on the Lie algebra of $\CG$ in order to contract the pairs of internal indices on expressions quadratic in $W^\alpha$.  This is the standard way in which supersymmetric Yang-Mills gauge theories in various spacetime dimensions are constructed in superspace \cite{west,wb}. 

Note the remarkable fact that our construction of the shift-vector sector in our topological gravity theory in Section~\ref{secgeofconn} in terms of the fields $N^i$, $S^i$ and $\bar S^I$, their superfield strengths $\CW$, and the corresponding constraints, takes precisely the form of the just reviewed standard supersymmetric Yang-Mills theory construction, with the following identifications:
\begin{itemize}
\item
  The underlying spacetime is the supertime, of dimension $(1|2)$, with coordinates $(t,\theta,\bar\theta)$.
\item
  The connections $\Gamma_M^\alpha$ are $N^i$ for $M=t$, $S^i$ for $M=\theta$, and $\bar S^i$ for $M=\bar\theta$, and with the adjoint index $\alpha$ being the multi-index $(i,x^k)$.  More precisely, in a language independent of the choice of coordinates on $\Sigma$, the Lie algebra $\CG$ is the infinite-dimensional algebra of vector fields on $\Sigma$.  Thus, the Lie algebra of this Yang-Mills theory is the Lie algebra of spatial diffeomorphisms $\CG=\diff(\Sigma)$!
\item
  One can check directly that the definitions of superfield strengths $\CW$ (\ref{eestrone}--\ref{eestrfive}) indeed correspond precisely to the Yang-Mills field strength definition in (\ref{eedefgw}). We recognize the structure constants of $\CG=\diff(\Sigma)$ in the expressions for $\CW$'s, and we also see that the first term in (\ref{eethird}) is the torsion term anticipated in (\ref{eedefgw}).
\item
  Our collection of constraints (\ref{ttbarw}) and (\ref{ttw}) is equivalent to the ``conventional constraints'' (\ref{eeconve}) of the standard superspace construction of supersymmetric Yang-Mills gauge theory.  
  \end{itemize}

Thus we reach a perhaps surprising conclusion:  The construction of the shift superfield sector in our topological quantum gravity on the superspace $\SM$ of dimension $(D+1|2)$ is precisely equivalent to the construction of supersymmetric Yang-Mills gauge theory on supertime of dimension $(1|2)$, and with the internal Lie algebra of gauge symmetries being the algebra of spatial diffeomorphisms $\diff(\Sigma)$ of the spatial slices $\Sigma$!

This intimate connection between conventional supersymmetric Yang-Mills theory with an internal symmetry $\CG$ on one hand, and the gauging of spatial diffeomorphisms in gravity on the other, can potentially be of some broader interest.  One is reminded of the BCJ color-kinematics duality \cite{bcj0,bcj,bcjre}, which relates amplitudes in Yang-Mills theories to amplitudes in gravity by replacing internal symmetry factors with kinematic factors.  It has been quite mysterious so far what kind of algebraic structure can underlie this procedure on the kinematic side.  Perhaps our relation between gravity and Yang-Mills may be useful in identifying the hidden algebraic structure on the side of kinematics, at the cost of singling out the role of time and making the description not manifestly relativistic.

While the parallel between our shift superfield sector and supersymmetric Yang-Mills is quite precise, there is one instance where the similarity stops:  Unlike compact finite Lie algebras $\CG$, our Lie algebra of spatial diffeomorphisms $\diff(\Sigma)$ does not have a constant invariant metric, and therefore one cannot construct standard quadratic kinetic terms for the action.  This is of course consistent with the prior knowledge that no such kinetic terms for the shift vector should exist.  In one wishes to construct an invariant metric on the Lie algebra of $\diff(\Sigma)$, it can only be done in a field-dependent way, by invoking a spatial metric $g_{ij}(x)$.  For two generators $\xi^i(x)$ and $\zeta^i(x)$ of $\diff(\Sigma)$, we define their inner product by
\be
(\xi,\zeta)=\int d^D x\,\sqrt{g}\,g_{ij}\,\xi^i\zeta^j.
\ee
This metric on $\diff(\Sigma)$ is thus field-dependent, and cannot be used to construct kinetic terms for the superfield strengths of the shift sector.  

\subsection{The action}

Using the covariant derivatives $\nabla_t G_{ij}, \CD G_{ij}$ and $\bar\CD G_{ij}$ constructed in the previous paragraphs,  the action for Type B theory can be easily constructed in the manifestly supersymmetric form in our $\CN=2$ superspace.

The kinetic term is covariantized to
\be
S_K=\int dt\,d^2\theta\,d^D x\,\sqrt{G}\left\{\left(\lambda_\perp G^{ik}G^{j\ell}-\lambda G^{ij}G^{k\ell}\right)\bar\CD G_{ij}\,\CD G_{k\ell}+\ldots\ \right\}.
\ee
(This kinetic term is valid in Type B theory; the corresponding kinetic terms in Type A and C cases are simply obtained by reducing $\CD$ or $\bar\CD$ to $\RD$ or $\barRD$ as appropriate.)  The superpotential stays the same as in our primitive theory, (\ref{eeactsw}).

The path integral for this theory is, in superspace language,%
\footnote{In the construction of the path-integral measure $\CD\mu$, one must keep in mind that $N^i, S^i$ and $\bar S^i$ are not independent, but related to each other by our constraints (\ref{sconst}) and (\ref{nssbar}).  At this stage, it might be better to switch from the superspace to the component formulation, in which the definition of the $\CN=2$ supersymmetric measure for the component fields is more straightforward.  Alternatively, Appendix~\ref{ssapp} solves the constraints and expresses our superfields in terms of unconstrained prepotential superfields.}
\be
\CZ=\int\SD\mu[G_{ij},N^i,S^i,\bar S^i]\,
\exp\left\{-\frac{1}{\hbar\kappa^2}(S_K-S_\CW)\right\}.
\label{eeactdiff}
\ee
This path integral requires further gauge fixing of the newly introduced spacetime $\diff (\Sigma)$ gauge symmetry, which we will discuss briefly in Section~\ref{sswzg}.  In addition, the same points that we presented in our brief discussion of the path integral of the primitive theory in Section~\ref{secactprim} apply here as well.

The main improvement compared to the primitive theory is that the flow equations for the metric $g_{ij}$ are now covariant under time-dependent spatial diffeomorphisms, with the path integral localizing to the solutions of
\be
\nabla_t g_{ij}=-\alpha_RR_{ij}+\frac{\alpha_R}{2}\left[1-\tilde\lambda(D-2)\right]g_{ij}R+\frac{\alpha_\Lambda}{2}g_{ij}.
\label{eelocdiff}
\ee
The possibility of modifying the Ricci flow by the time-dependent spatial diffeomorphism generated by $n^i$ has been very useful in the mathematical theory, where it is known as ``DeTurck's trick''.  We are now in a position to see how these techniques emerge in our quantum gravity, as a part of the process of gauge fixing the spatial diffeomorphism symmetry.

\subsection{Wess-Zumino gauge}
\label{sswzg}

In order to construct a theory with spatial diffeomorphism gauge symmetry generated by bosonic generators $\xi^i(t,x^k)$ in a way manifestly consistent with $\CN=2$ global supersymmetry of supertime, we first promoted $\xi^i$ to a superfield of symmetry generators $\Xi^i$, and used superspace techniques to find a theory invariant under the much larger symmetry generated by $\Xi^i$.  Now we need to decide how to interpret -- or eliminate -- the additional gauge symmetries contained in $\Xi^i$ of (\ref{eexisym}), \ie, symmetries generated by
\be
\zeta^i,\qquad \eta^i,\quad\textrm{and}\quad \alpha^i,
\label{eewzg}
\ee
so that we reduce the gauge group back to the desired $\diff(\Sigma)$.  Until this point, our strategy for the gauging process has closely parelleled the construction of supersymmetric relativistic gauge theories (see, \eg, \cite{west} for a review and introduction), which also offers a natural way of reducing the gauge symmetries to the bosonic ones, known as Wess-Zumino gauge \cite{wessz}:  One simply sets all the higher components of the gauge superfield, in our case the shift superfield $N^i$, to zero.   This is algebraically possible, leads to no additional constraints, and leaves only the bosonic $\xi^i$ symmetry unfixed.   

We will adopt this Wess-Zumino gauge for the spatial diffeomorphism symmetry in our theory.  Thus, the action that appears in (\ref{eeactdiff}) in Wess-Zumino gauge remains gauge invariant only under the bosonic $\diff(\Sigma)$ symmetry, but still in a way consistent with the $\CN=2$ supersymmetry.  The path-integral measure is also correspondingly reduced in Wess-Zumino gauge.

Having disposed of the higher gauge symmetries (\ref{eewzg}), we must next decide how to treat the remaining bosonic gauge symmetries $\diff(\Sigma)$.  There are several useful options.  First, we can leave the theory in its manifestly $\diff(\Sigma)$ invariant form for as long as possible, and introduce its gauge fixing by the standard Faddeev-Popov ghosts when necessary (for example, for developing Feynman diagrams around a given background).  This ``equivariant'' approach is the strategy often preferred in topological field theories.  It is followed for example for relativistic topological Yang-Mills, where the Yang-Mills gauge symmetry is typically left unfixed.

Altenatively, there might be reasons why one may want to fix, fully or partially, the $\diff(\Sigma)$ symmetry.  For the topological gravity of the Ricci flow, this option turns out to be very useful for the comparison to the mathematical literature.  In fact, we will see three different natural gauge choices, each corresponding to an operation performed in the mathematical theory of the Ricci flow.  We refer to them as ``DeTurck gauge,'' ``Perelman gauge'', and ``Hamilton gauge''.

\begin{itemize}
\item
\textit{Perelman gauge:}  In this gauge, one simply sets the shift vector to zero,
\be
n^i(t,x^j)=0.
\label{eehgauge}
\ee
In the context of gravity and spatial diffeomorphism symmetry, this is the analog of temporal gauge. Adopting this gauge choice, the covariant time derivative $\nabla_t g_{ij}$ in the localization equation (\ref{eelocdiff}) is reduced to the ordinary time derivative $\dot g_{ij}$, as in the original form (\ref{eehamrf}) of Hamilton's Ricci flow (which was not invariant under time-dependent spatial diffeomorphisms).  

\item
  \textit{Hamilton gauge:}  If the theory contains another field $h(t,x^i)$, which transforms as a scalar under spatial diffeomorphisms, one can replace (\ref{eehgauge}) with 
  \be
  n^i(t,x^j)=g^{ik}\p_k h.
  \label{eepgauge}
  \ee
We do not have any such scalars in the theory yet, but will see that this type of gauge will be useful when we extend the gauge symmetries to foliation-preserving diffeomorphisms of spacetime.  The gauge-fixing condition (\ref{eepgauge}) played an important role in Perelman's original approach to Ricci flow, in particular in re-establishing the relation to the original Hamilton-Ricci flow (\ref{eehamrf}); the role of $h$ was played by Perelman's dilaton field $\phi$.
  
\item
  \textit{DeTurck gauge:}  This is the context in which the original DeTurck trick first appeared \cite{det}.  To define this gauge, one first chooses a fixed fiducial metric $\tilde g_{ij}$ on $\Sigma$, and sets 
\be
n^i=g^{jk}\left(\Gamma^i_{jk}-\tilde\Gamma^i_{jk}\right),
\ee
where $\Gamma^i_{jk}$ and $\tilde \Gamma^i_{jk}$ are the Christoffel symbols representing the torsion-free Levi-Civit\`a connections of $g_{ij}$ and $\tilde g_{ij}$ respectively (see, \eg , Ch.~3.3 of \cite{rfi} or Ch.~2.6 of \cite{hrf} for additional mathematical context and motivation).  This choice obviously breaks spatial diffeomorphism invariance.  The usefulness of this gauge choice stems from the fact that the Ricci flow equation in this gauge is found to be manifestly parabolic, a property not obvious in other gauges, and definitely untrue for the gauge-unfixed flow equation (\ref{eehamrf}) (which is parabolic only modulo spatial diffeomorphisms, or ``weakly'' parabolic).  In turn, this manifest parabolicity leads to a simple proof of the existence and uniqueness theorem, stating that a solution of the initial value problem for the flow equation exists for some amount of time $\varepsilon>0$, and that on that time interval the solution is unique.  
  
\end{itemize}

The first two of these gauge choices are going to be particularly useful once we extend the gauge symmetries to foliation-preserving spacetime diffeomorphisms, especially in the theory with the nonprojectable lapse.  

\section{The gauge theory:  Gauging time translations}
\label{sstime}

Next, we wish to gauge time translations, or at least those that preserve the preferred foliation of spacetime.  In the bosonic theory, such foliation-preserving time diffeomorphisms are generated by
\be
\delta t =f(t).
\ee
To promote them to a gauge symmetry, we introduce a new field, the lapse function $n(t)$, which transforms as
\be
\delta n=f\dot n+\dot fn.
\ee
Multiplying the covariant time derivative $\nabla_tg_{ij}$ with the inverse lapse function, we obtain $(1/n)\nabla_tg_{ij}$, which transforms as a scalar under time diffeomorphisms.  Such scalars can then be used to build invariant Lagrangians, which take the form of the covariant spacetime volume element
\be
d\CV (g,n)\equiv d^{D}x\,dt\,n\sqrt{g},
\ee
mutliplied by any scalar function made out of the available ingredients.

We will now generalize this gauging procedure to our supersymmetric case.  In order to make $f(t)$ consistent with supersymmetry, it must first be promoted into a superfield,
\be
F(t,\theta,\bar\theta)=f(t)+\theta\varphi(t)+\bar\theta\bar\varphi(t)+\theta\bar\theta \gamma(t).
\label{eetimef}
\ee
In addition, we may choose this superfield to be further constrained, for example by a chirality condition.  Note that the superfield of time reparametrizations is independent of $x^i$, reflecting the fact that our gauge symmetries preserve the structure of the spacetime foliation $\SM_\CF$ by spatial manifolds $\Sigma$ of constant $t,\theta,\bar\theta$.  

\subsection{The projectable case}

In this section, we will construct the minimal theory consistent with the gauge symmetries of (\ref{eetimef}).  This theory will have a projectable lapse $n(t)$, promoted to a superfield $N(t)$.  As in the case of spatial diffeomorphisms, there are three versions of the theory, depending on whether we impose a chirality or antichirality condition on $N$, or keep the superfield unconstrained.

\subsubsection{Type C theory}

We begin with our chiral Type C theory of Section~\ref{sectypec}, and we extend the gauge symmetry of the chiral spatial diffeomorphisms $\Xi^i(t,\theta)$ to also include the chiral version of time reparametrization symmetry generated by $F$ which satisfies $\barRD F=0$.

The gauge transformations of the previously introduced superfields $G_{ij}, N^i$ and $S^i$ are:
\bea
\delta G_{ij}&=&F\dot G_{ij}+\Xi^k\p_kG_{ij}+\p_i\Xi^kG_{kj}+\p_j\Xi^kG_{ik},\\
\delta N^i&=&F\dot N^i-\dot FN^i+\dot\Xi^i+\Xi^k\p_kN^i-\p_k\Xi^iN^k,\\
\delta S^i&=&F\dot S^i-\RD F\barRD S^i+\RD\Xi^i+\Xi^k\p_kS^i-\p_k\Xi^iS^k.
\eea
The first two of these rules follow straightforwardly from the requirement that the bosonic component fields $g_{ij}$ and $n^i$ transform under $f(t)$ as in the bosonic theory.  The third rule follows from the requirement that the constraint (\ref{constns}) that relates $N^i$ to $S^i$ be preserved under the time reparametrizations.

In order to construct the theory with $F$ gauge symmetry, we could introduce a supervielbein on $(t,\theta,\bar\theta)$ (which would be a $3\times 3$ matrix of superfields) and impose enough constraints on it so that we reduce the number of independent component fields to the bosonic lapse function and its superpartner under $Q$.  Here we will follow a much more straightforward ``bottom-up'' strategy, and will return to the supervielbein interpretation below once our construction is complete.

Consider the derivatives $\nabla_tG_{ij}$, $\CD G_{ij}$ and $\barRD G_{ij}$, which serve as ingredients for building our Lagrangian in superspace.  Under $F$, some of these derivatives do not transform as scalars.  The task is to modify them minimally so that the modified derivatives do transform as scalars under $F$, and can then be again used as simple ingredients for constructing gauge invariant Lagrangians.  

Start with the time derivative $\nabla_t$.  In the bosonic theory, its covariantization under $f(t)$ is simply accomplished by multiplying it with the inverse lapse function $1/n$.  In the supersymmetric case, we introduce superfield $E(t,\theta,\bar\theta)$ whose lowest component is $1/n$, and observe that $E\nabla_tG_{ij}$ transforms as a scalar under $F$,
\be
\delta (E\nabla_tG_{ij})=F\p_t(E\nabla_tG_{ij}),
\ee
if we postulate that $E$ transform as
\be
\delta E=F\dot E-\dot F E.
\ee

Our next step is to covariantize similarly the remaining derivatives $\CD G_{ij}$ and $\barRD G_{ij}$.  Since in Theory C, $\barRD G_{ij}$ does not contain a gauge field, it transforms as a scalar under $F$.  On the other hand, $\CD G_{ij}$ contains $S^i$ terms and does not transform as a scalar, and therefore requires a modification.  As in the case of the time derivative, the first step is to introduce a new superfield $\CE(t,\theta,\bar\theta)$ and replace $\CD G_{ij}$ with $\CE\CD G_{ij}$.  This by itself is not sufficient, since the transformation of $\CD G_{ij}$ under $F$ will also contain terms proportional to $\nabla_tG_{ij}$.  One must introduce one additional, odd superfield $\Theta$, and shift $\CE\CD G_{ij}$ by an additive term $\Theta\nabla_tG_{ij}$.  Postulating the transformation rules
\bea
\delta\CE&=&F\dot\CE,\\
\delta\Theta&=&-\CE\,\RD F+F\dot\Theta-\dot F\Theta
\eea
then ensures that
\be
\CE\,\CD G_{ij}+\Theta\nabla_tG_{ij}
\ee
transforms as a scalar under $F$.

Thus, the covariantization of the derivatives consistently with supersymmetry requires the introduction of three superfields $E,\Theta$ and $\CE$, which play the role which in the bosonic theory was played by the (inverse) lapse function.  Clearly, these three superfields must be further constrained, so that they do not lead to a proliferation of gauge-invariant component fields for which we have no interpretation or desire.  

The first such constraint is easy to propose:  Since $\CE$ transforms under $F$ covariantly as a scalar, it is consistent to set
\be
\label{constce}
\CE=1.
\ee

Then there must be a constraint that relates $E$ to $\Theta$.  A closer examination of the transformation properties reveals that $1-\barRD\Theta$ transforms the same way as $E$, and we therefore impose the constraint
\be
\label{consteth}
E=1-\barRD\Theta.
\ee
Note that this constraint implies that $E$ is chiral, and therefore only contains two components, as it should:  The inverse lapse function and its superpartner under $Q$.

Finally, $\Theta$ also must satisfy a constraint which reduces its components to two.  Much like the gauge field $S^i$ already present in Type C theory, $\Theta$ should satisfy some covariantized version of the antichirality constraint; the unique combination that transforms correctly is
\be
\label{constthth}
\RD\Theta=-\Theta\dot\Theta.
\ee
This completes our construction of the projectable version of Theory C with chiral $F(t,\theta)$ and $\Xi^i(t,\theta)$ gauge symmetries.

\subsubsection{Geometric interpretation of constraints: Flatness of supertime}

Before considering the lift to the nonprojectable case, we make one additional observation, which will be useful in the more complicated cases below.  The constraints postulated on $E$ and $\Theta$ above have a very natural geometric interpretation -- they simply state that the vielbein geometry of supertime is flat!  Indeed, it is natural to consider the graded commutators of the covariantized derivatives
\be
\label{ders}
\SD_t\equiv E\nabla_t,\qquad \SD_\theta\equiv\CE\CD+\Theta\nabla_t,\qquad \SD_{\bar\theta}\equiv\barRD
\ee
acting on $G_{ij}$.  The (super)curvature of the geometry on supertime represented by our superfielbein fields $E,\CE$ and $\Theta$ is then defined as the deviation of such graded commutators from the standard graded commutation relations satisfied before the gauging of time translations by $\nabla_t$, $\CD$ and $\barRD$.  (Recall that the latter three operators already represent a flat Yang-Mills connection of $\diff(\Sigma)$, as we found out in Section~\ref{secb}.)

A straightforward evaluation of all the graded commutators of (\ref{ders}) shows that our constraints (\ref{consteth}), (\ref{constthth}) (together with (\ref{constce})) imply the vanishing of all the curvature terms.  

\subsection{The nonprojectable case}

In bosonic nonrelativistic gravity of the Lifshitz type, the more interesting and useful theory is obtained when the lapse function is allowed to be nonprojectable, $n(t,x^i)$.
It is then natural to ask whether the vielbein superfields $E$, $\CE$ and $\Theta$ can be promoted into spacetime fields, \ie , allowed to depend on $x^i$.%
\footnote{In the geometry of foliations, and in the literature on the bosonic version of nonrelativistic gravity, such fields are commonly referred to as ``nonprojectable'', in contrast to the ``projectable'' fields which are functions of only the leaves of the foliation, and whose lift to all of spacetime is simply via the pull-back by the natural projection of the foliation.}
Note that the gauge symmetries will stay the same foliation-preserving diffeomorphisms of spacetime as in our previous construction with the projectable lapse superfields; in particular, the generator of time reparametrizations $F$ is only a function of $t$ and/or $\theta,\bar\theta$.

\subsubsection{Type C theory}

In this section, we will show that such a nonprojectable version of our theory does indeed exist, first in the Type C case.

When our lapse-sector superfields $E,\CE$ and $\Theta$ are extended to be nonprojectable superfields on spacetime, they transform as scalars under $\Xi^i$.  Thus, their full transformation rules are 
\bea
\delta E&=&F\dot E-\dot F E+\Xi^k\p_k E,\\
\delta\Theta&=&-\CE\,\RD F+F\dot\Theta-\dot F\Theta+\Xi^k\p_k\Theta,\\
\delta \CE&=&F\dot\CE+\Xi^k\p_k\CE.
\eea

What is the nonprojectable version of the constraints?  Consider first the superfield $\CE$.  Since it transforms as a scalar under both $F$ and $\Xi^i$, it is again consistent to set it equal to a constant, which we choose without any loss of generality to be $\CE=1$.  We will impose this constraint from now on, and return to the more general case of arbitrary $\CE$ later, in Section~\ref{secgeoe}.

The constraint relating $\Theta$ and $E$ stays the same as in the projectable case,
\be
E=1-\barRD\Theta,
\ee
but the constraint on $\Theta$ is modified to
\be
\RD\Theta-S^k\p_k\Theta=-\Theta(\dot\Theta-N^k\p_k\Theta).
\ee
The easiest way to derive these constraints is to evaluate again all the conditions for the vanishing of the supertime curvatures of the nonprojectable fields $E$ and $\Theta$, as we discussed in the projectable version above.  

\subsubsection{Type B theory}
\label{sectypebnon}

Now we extend the gauging of time translations to our balanced Type B theory, in which $F$ is an unconstrained superfield.  We jump directly to the nonprojectable case; the projectable one results by simply restricting the lapse superfields to be independent of $x^i$.  Similarly, the gauging of time translations in the antichiral Type A theory will follow by restricting $F$ to be antichiral, and the lapse superfields correspondingly constrained as well; see Section~\ref{sectae}.

The gauge parameter $F(t,\theta,\bar\theta)$ is of course independent of $x^i$, but otherwise unconstrained.  We introduce superfields $E, \CE,\bar\CE,\Theta$ and $\bar\Theta$ to covariantize all derivatives.  The transformation rules of all fields under the spacetime gauge symmetries are:
\bea
\label{fulltrsrl}
\delta G_{ij}&=&F\dot G_{ij}+\Xi^k\p_kG_{ij}+\p_i\Xi^kG_{kj}+\p_j\Xi^kG_{ik},\nonumber\\
\delta N^i&=&F\dot N^i-\dot FN^i+\dot\Xi^i+\Xi^k\p_kN^i-\p_k\Xi^iN^k,\nonumber\\
\delta S^i&=&F\dot S^i+\RD F\,N^i+\RD\Xi^i+\Xi^k\p_kS^i-\p_k\Xi^iS^k,\nonumber\\
\delta{\bar S}^i&=&F\dot{\bar S}^i+\barRD F\,N^i+\barRD\Xi^i+\Xi^k\p_k\bar S^i-\p_k\Xi^i\bar S^k,\nonumber\\
\delta E&=&F\dot E-\dot F E+\Xi^k\p_k E,\\
\delta\Theta&=&-\CE\,\RD F+F\dot\Theta-\dot F\Theta+\Xi^k\p_k\Theta,\nonumber\\
\delta\bar\Theta&=&-\bar\CE\,\barRD F+F\dot{\bar\Theta}-\dot F\bar\Theta+\Xi^k\p_k\bar\Theta,\nonumber\\
\delta \CE&=&F\dot\CE+\Xi^k\p_k\CE,\nonumber\\
\delta \bar\CE&=&F\dot{\bar\CE}+\Xi^k\p_k\bar\CE.\nonumber
\eea
With these rules, the following derivatives transform as scalars:
\be
\label{scovders}
\SD_tG_{ij}\equiv E\nabla_tG_{ij},\qquad\SD_\theta G_{ij}\equiv\CE\CD G_{ij}+\Theta\nabla_tG_{ij},\qquad\SD_{\bar\theta}G_{ij}\equiv\bar\CE\bar\CD G_{ij}+\bar\Theta\nabla_tG_{ij}.
\ee
As in the simpler Type C case, the supervielbein fields $E, \CE,\bar\CE,\Theta$ and $\bar\Theta$ must satisfy a number of constraints.  First, we will follow our strategy from Type C theory and set
\be
\label{econ1}
\CE=1,\qquad\bar\CE=1.
\ee
$E$ is then constrained to be expressed in terms of $\Theta$ and $\bar\Theta$ and their derivatives,
\be
\label{econ2}
E=1-\RD\bar\Theta+S^k\p_k\bar\Theta-\barRD\Theta+\bar S^k\p_k\Theta-\Theta(\dot{\bar\Theta}-N^k\p_k\bar\Theta)-\bar\Theta(\dot\Theta-N^k\p_k\Theta).
\ee
Finally, $\Theta$ and $\bar\Theta$ are constrained to satisfy
\bea
\label{econ3}
\RD\Theta-S^k\p_k\Theta&=&-\Theta(\dot\Theta-N^k\p_k\Theta),\\
\label{econ4}
\barRD\bar\Theta-\bar S^k\p_k\bar\Theta&=&-\bar\Theta(\dot{\bar\Theta}-N^k\p_k\bar\Theta).
\eea
These constraints again leave the desired number of four independent component fields:  the inverse lapse function and its superpartners under $Q$ and $\bar Q$.

\subsubsection{Constraints as the flatness of supertime}
\label{secgeoe}

It is instructive to check the geometric origin of our constraints in the nonprojectable Type B theory, which we will again interpret simply as the statement of the flatness of our supervielbein fields on supertime.  We also take this opportunity to address our earlier somewhat {\it ad hoc\/} step of setting $\CE=\bar\CE=1$, and will allow these superfields now to be unconstrained.  Thus, our covariant derivatives are those we constructed in (\ref{scovders}), before imposing any \textit{ad hoc} constraints.

Now we evaluate their graded commutators to evaluate the conditions for flatness.  We begin by evaluating
\bea
\left\{\SD_\theta,\SD_\theta\right\}G_{ij}=2(\CE\CD+\Theta\nabla_t)^2G_{ij}&=&2\left[\CE (\RD\CE-S^k\p_k\CE)+\Theta(\dot\CE-N^k\p_k\CE)\right]\CD G_{ij}\\
&+&2\left[\CE (\RD\Theta-S^k\p_k\Theta)+\Theta(\dot\Theta-N^k\p_k\Theta)\right]\nabla_t G_{ij}.
\eea
The vanishing of the corresponding curvature requres that the right-hand side be zero, which implies the constraints
\bea
\CE (\RD\CE-S^k\p_k\CE)+\Theta(\dot\CE-N^k\p_k\CE)&=&0,\\
\CE (\RD\Theta-S^k\p_k\Theta)+\Theta(\dot\Theta-N^k\p_k\Theta)&=&0.
\eea
Similarly, the anticommutator $\left\{\SD_{\bar\theta},\SD_{\bar\theta}\right\} G_{ij}$ gives the analogous condition for the barred quantities,
\bea
\bar\CE (\barRD\bar\CE-\bar S^k\p_k\bar\CE)+\bar\Theta(\dot{\bar \CE}-N^k\p_k\bar\CE)&=&0,\\
\bar\CE (\barRD\bar \Theta-\bar S^k\p_k\bar \Theta)+\bar\Theta(\dot{\bar\Theta}-N^k\p_k\bar\Theta)&=&0.
\eea
Next, we evaluate the anticommutator
\bea
&&\left\{\SD_\theta,\SD_{\bar\theta}\right\}G_{ij}=\left[\CE(\RD\bar\CE-S^k\p_k\bar\CE)+\Theta(\dot{\bar\CE}-N^k\p_k\bar\CE)\right]\bar\CD G_{ij}\\
&&\qquad\qquad\qquad
\qquad{}+\left[\bar\CE(\bar \RD\CE-\bar S^k\p_k\CE)+\bar\Theta(\dot\CE-N^k\p_k\CE)\right]\CD G_{ij}\nonumber\\
&+&\left[-\CE\bar\CE+\CE(\RD\bar\Theta-S^k\p_k\bar\Theta)+\bar\CE(\barRD\Theta-\bar S^k\p_k\Theta)+
\Theta(\dot{\bar\Theta}-N^k\p_k\bar\Theta)+\bar\Theta(\dot\Theta-N^k\p_k\Theta)\right]\nabla_tG_{ij}.\nonumber
\eea
The flatness condition then requires that this anticommutator be equal to $-E\nabla_tG_{ij}$, implying the following constraints:
\bea
\CE(\RD\bar\CE-S^k\p_k\bar\CE)+\Theta(\dot{\bar\CE}-N^k\p_k\bar\CE)&=&0,\\
\bar\CE(\barRD\CE-\bar S^k\p_k\CE)+\bar\Theta(\dot\CE-N^k\p_k\CE)&=&0,\\
E=\CE\bar\CE-\CE(\RD\bar\Theta-S^k\p_k\bar\Theta)-\bar\CE(\barRD\Theta-\bar S^k\p_k\Theta)&-&\Theta(\dot{\bar\Theta}-N^k\p_k\bar\Theta)-\bar\Theta(\dot\Theta-N^k\p_k\Theta).\ \ \ \ 
\eea
Finally, the commutators of the odd superderivatives with $E\nabla_t$ imply that the remaining conditions of vanishing curvature are
\bea
E(\dot\CE-N^k\p_k\CE)&=&0,\\
E(\dot\Theta-N^k\p_k\Theta)&=&\CE(\RD E-S^k\p_kE)+\Theta(\dot E-N^k\p_kE),
\eea
and
\bea
E(\dot{\bar\CE}-N^k\p_k\bar\CE)&=&0,\\
E(\dot{\bar\Theta}-N^k\p_k\bar\Theta)&=&\bar\CE(\barRD E-\bar S^k\p_kE)+\bar\Theta(\dot E-N^k\p_kE).
\eea
Note that in the projectable case, assuming that $E,\CE$ and $\bar\CE$ are invertible, the constraints imply that $\CE$ and $\bar\CE$ are constants, which in retrospect justifies our choice of setting them equal to 1 from the outset.  In the nonprojectable case, the constraints are solved by $\CE$ and $\bar\CE$ whose covariant time derivative is zero.

It is easy to check that our constraints (\ref{econ1}--\ref{econ4}) represent the minimal set of constraints which imply all the other conditions of flatness of supertime by Bianchi identities.  The constraints can be solved explicitly by finding the component expressions for the lapse-sector superfields, repeating the steps we took in Section~\ref{secgeofconn} when we solved the analogous constraints in the shift sector.

\subsubsection{Type A theory}
\label{sectae}

The antichiral Type A theory results by restricting $F$ and $\Xi^i$ to be antichiral:
\be
\RD F=0,\qquad \RD\Xi^i=0,
\ee
and by setting
\be
\CE=1,\qquad \Theta=0,\quad \mathrm{and}\quad S^i=0
\ee
in the rules specified in the ``umbrella'' theory of Type B presented above.  We will not require any further details about this Type A theory in the rest of this paper.

\subsection{The supervielbein approach}

Now we are ready to compare our approach to the top-down construction using supervielbeins.  (We present the construction only for Type B theory, and for simplicity for its projectable version, with the projectable Type A and C cases following by a simple reduction.)

In the supervielbein approach, one postulates the existence of a $3\times 3$ supervielbein matrix of superfields
\be
e_M{}^A,\qquad M\in\{t,\theta,\bar\theta\},\quad A\in\{0,\vartheta,\bar\vartheta\},
\ee
where $M$ is the coordinate index and $A$ is the internal tangent-space index on supertime, and find enough constraints on $e_M{}^A$ to reduce them drastically to just four component fields: the lapse function and its superpartners.

Here we will establish the connection between our bottom-up construction involving superfields $E,\CE,\bar\CE,\Theta$ and $\bar\Theta$, and the full top-down supervielbein construction.  It will be more convenient for us to work with the inverse supervielbein $e_A{}^M$, which is simply defined to interpolate between the coordinate basis $\p_M$ in the tangent space to supertime, and the moving-frame basis $D_A$ whose three elements are labeled by the internal index $A$:
\be
D_A=e_A{}^M\p_M.
\ee
On the rigid supertime before the introduction of $E$, we have $D_A=(\p_t,\RD,\barRD)$, and the standard flat (inverse) supervielbein is given by
\be
e_A^{(0)}{}^M=
\begin{pmatrix}
  1&&0&&&0&\\
  -\bar\theta&&1&&&0&\\
  0&&0&&&1&
\end{pmatrix}.
\ee

Once we gauge time translations, the $D_A$'s are given by the covariantized derivatives $E\p_t,\CE\RD+\Theta\p_t$ and $\bar\CE\bar\CD+\bar\Theta\p_t$, and the inverse supervielbein becomes
\be
\label{curvede}
e_A{}^M=
\begin{pmatrix}
  E&&0&&&0&\\
  \Theta-\bar\theta\CE&&\CE&&&0&\\
  \bar\Theta&&0&&&\bar\CE&
\end{pmatrix}.
\ee
Consider the generic superdiffeomorphism of supertime, generated in our coordinates $\tau^M\equiv(t,\theta,\bar\theta)$ by some
\be
\delta\tau^M=F^M(\tau^N).  
\ee
Under this superdiffeomorphism, the supervielbein transforms geometrically, as
\be
\label{trsfme}
\delta e_M{}^A=F^N\p_N e_M{}^A+\p_MF^Ne_N{}^A,
\ee
and analogously for the inverse supervielbein $e_A{}^M$.

Our gauge symmetry of gauged time translations is a specific subalgebra of this, consisting only of supertime-dependent time reparametrizations:  $F^t=F(t,\theta,\bar\theta)$, and $F^\theta=F^{\bar\theta}=0$.  Therefore, we should verify that our constrained supervielbein (\ref{curvede}) which we derived in our bottom-up approach indeed transforms under $F$ according to (\ref{trsfme}).   It is a pleasing check that with our transformation rules for $E,\Theta,\bar\Theta,\CE$ and $\bar\CE$ established above, the vielbein indeed transforms geometrically as anticipated.  For example, the variation $\delta e_\vartheta{}^t$ implied by (\ref{trsfme}) should be
\be
\delta e_\vartheta{}^t=F\,\dot e_\vartheta{}^t-\dot F\,e_\vartheta{}^t-e_\vartheta{}^\theta\p_\theta F.
\ee
Substituting from (\ref{curvede}), this predicts
\bea
\delta(\Theta-\bar\theta\CE)&=&F(\dot\Theta-\bar\theta\dot\CE)-\dot F(\Theta-\bar\theta\CE)-\CE\p_\theta F\nonumber\\
&&{}=F(\dot\Theta-\bar\theta\dot\CE)-\dot F\Theta-\CE\,\RD F,
\eea
which exactly matches the result obtained directly by using the projectable version of the transformation rules (\ref{fulltrsrl}).

We note that geometrically, the gauge symmetries that we have implemented on our system are those of spacetime diffeomorphisms that preserve the structure of a nested double foliation of the spacetime supermanifold,
\be
\SM\rightarrow\SM_0^{1|2}\rightarrow\SM_0^{0|2}.
\ee
In particular, the supertime $\SM_0^{1|2}$ itself is naturally foliated by leaves of constant $(\theta,\bar\theta)$, with the leaves parametrized by $t$.  

\subsection{The action}

The covariant volume element on $\SM$ is now
\be
d\CV (G,E)=dt\,d^2\theta\,d^D x\,\frac{\sqrt{G}}{E}.
\ee
In the projectable theory, the lowest-dimension kinetic term invariant under our full spacetime gauge symmetry is given by
\be
\label{eeactskfdif}
S_K=\int dt\,d^2\theta\,d^D x\,\frac{\sqrt{G}}{E}\left\{(\lambda_\perp G^{ik}G^{j\ell}-\lambda G^{ij}G^{k\ell})\SD_{\bar\theta} G_{ij}\,\SD_\theta G_{k\ell}+\ldots\ \right\};
\ee
as in the primitive theory, we again set $\lambda_\perp=1$ for simplicity.  On the other hand, the superpotential now allows for a more refined structure.

\subsubsection{The superpotential and Perelman's $\CF$-functional}

In the bosonic nonprojectable gravity of the Lifshitz type, it is well appreciated that new ingredients appear and can be used to construct new terms in the action.  In particular, the spatial derivatives $\p_i n$ of the nonprojectable lapse transform as a spatial one-form, and it can give rise to new invariant Lagrangian terms.  In our $\CN=2$ supersymmetric theory, we similarly find new ingredients, which give rise to new invariants that can appear in the superpotential.  In particular, 
\be
A_i\equiv \frac{\p_iE}{E}
\ee
transforms as a spatial one-form and a time scalar,
\be
\delta A_i=F\dot A_i+\Xi^k\p_kA_i+\p_i\Xi^kA_k.
\ee
We can form new invariants in the action, made of the appropriate contractions of $A_i$. In terms of the superfield $\Phi$ defined via
\be
\Phi\equiv\log E,
\ee
we simply have $A_i=\p_i\Phi$.  The superpotential part of the action is now
\be
S_\CW=\int dt\,d^2\theta\,d^D x\,e^{-\Phi}\sqrt{G}\left\{\alpha_R R^{(G)}+\alpha_\Phi\,G^{ij}\p_i\Phi\p_j\Phi+\alpha_\Lambda\right\},
\label{eenonpw}
\ee
for some coupling constants $\alpha_R$, $\alpha_\Phi$ and $\alpha_\Lambda$.  We recognize $S_\CW$ as a superfield version of Perelman's $\CF$-functional (\ref{eeperlf}), simply generalized to include the cosmological constant term!  Note that the role of Perelman's ``dilaton'' is played in our theory by the logarithm of the nonprojectable lapse function.  These two results are the central results of the present paper.

\subsubsection{Localization equations and generalizations of Perelman's Ricci flow}
\label{ssgauge}

This picture can be fleshed out even more by switching to the component formulation.  As in Section~\ref{sshamloc}, we will again suppress all the fermionic terms which are uniquely determined from supersymmetry, and focus only on the bosonic fields.  In addition, for reasons of simplicity, we present the results only for Type~C or Type~A theory.

The bosonic component action corresponding to the superspace action (\ref{eeactskfdif}) and (\ref{eenonpw}) is:
\bea
S_{\mathrm{bose}}&=&-\frac{1}{\kappa^2}\int dt\,d^Dx\,\sqrt{g}n \left(g^{ik}g^{j\ell}-\lambda g^{ij}g^{k\ell}\right) B_{ij} B_{k\ell}\nonumber\\
&&\qquad\qquad\qquad\qquad {}+\frac{1}{\kappa^2}\int dt\,d^Dx\,\sqrt{g}\left(g^{ik}g^{j\ell}-\lambda g^{ij}g^{k\ell}\right) B_{ij}\,\nabla_tg_{k\ell}\nonumber\\
&&\quad-\frac{1}{\kappa^2}\int dt\,d^Dx\,\sqrt{g}n\,B_{ij}\left\{
\alpha_R\left(\frac{1}{2}Rg^{ij}-R^{ij}\right)+\left(\frac{1}{2}\alpha_\Phi-\alpha_R\right)g^{ij}(\p\phi)^2\right.\nonumber\\
&&\quad\left.{}+\alpha_R\,g^{ij}\,\Delta\phi+(\alpha_R-\alpha_\Phi)g^{ik}\,g^{j\ell}\,\p_k\phi\,\p_\ell\phi-\alpha_R\,g^{ik}\,g^{j\ell}\,\nabla_k\p_\ell\phi+\frac{1}{2}\alpha_\Lambda g^{ij}\right\}.
\eea
The saddle points of the action correspond to the spatial metric $g_{ij}$  satisfying the appropriate flow equation, governed by the variation of a functional which is a direct generalization of Perelman's $\CF$-functional.  Integrating out the bosonic auxiliary field $B_{ij}$ we obtain the localization equations, in the form of a flow equation covariantized with respect to foliation-preserving spacetime diffeomorphisms,
\bea
\frac{1}{n}\left(\dot g_{ij}-\nabla_in_j-\nabla_jn_i\right)&=&-\alpha_R\,R_{ij}+\frac{\alpha_R}{2}\left[1-\tilde\lambda(D-2)\right]g_{ij}R\nonumber\\
&+&(\alpha_R-\alpha_\Phi)\p_i\phi\p_j\phi-\left[\left(\alpha_R-\frac{\alpha_\Phi}{2}\right)(1-\tilde\lambda D)+(\alpha_R-\alpha_\Phi)\tilde\lambda\right]g_{ij}(\p\phi)^2
\nonumber\\
&&\qquad{}+\alpha_R\left[1-\tilde\lambda(D-1)\right]g_{ij}\Delta\phi-\alpha_R\nabla_j\p_j\phi+\frac{1}{2}\alpha_\Lambda\,g_{ij}.
\label{eeforest}
\eea
This is a multi-parameter family of generalized Ricci-type flow equations for the spatial metric $g_{ij}$.

In contrast to $g_{ij}$, the lapse field $n=\exp(-\phi)$ does not yet receive any nontrivial time evolution from localization.  In Type~A or Type~C theory, this is because the chirality condition on $N$ eliminates the auxiliary field associated with $n$, and the topological symmetries of the theory are not yet fully gauge-fixed.  Even in Type B theory, however, the required lowest-dimension kinetic term for $n$ (or $\phi$) cannot appear.  This is simply because our spacetime foliation-preserving gauge invariance, which has so far been unfixed, prevents such terms from being gauge invariant.  This is as far as the gauge-invariant theory can take us, and to make a closer contact with the exact form of Perelman's flow, additional gauge fixing steps will be necessary.

\subsubsection{Physical versus topological theory}
\label{ssreal}

We return to the possibility of analytically continuing the topological theory from imaginary time to real time, raised briefly in our comments on the path integral (\ref{pathintp}) of the primitive theory.

In the case of  relativistic quantum field theories, such a direct continuation of a topological field theory to real time would have little sense:  In real time, the fermions would violate the spin-statistics theorem, and the field theory could not be interpreted as a unitary theory of propagating degrees of freedom, at least not without some additional difficult ``untwisting'' steps.  In contrast, in the case of topological nonrelativistic gravity, one can at least entertain the possibility of continuing the theory to real time and interpreting it as a theory with propagating degrees of freedom.  This would require an analytic continuation of our superspace, such that $\bar\theta$ and $\theta$ would now be complex, and conjugates of each other.  This is needed so that the component fields could have physically sensible dispersion relations at least in some portions of the space of the coupling constants $\lambda$ and $\alpha$, and their quanta could be interpreted as physical particles.  Since there is no spin-statistics theorem in nonrelativistic field theory, this continuation could in principle lead to a consistent nonrelativistic gravity with gravitons and their superpartners with $\CN=2$ supersymmetry.  The absence of the spin-statistics theorem in nonrelativistic systems makes the boundary between Faddeev-Popov ghosts and propagating physical fields interestingly fuzzy, and the appealing direct relation between a topological and a physical theory possible in principle.  

However, before making sense of this rotation to real time and a nonrelativistic gravity with propagating degrees of freedom, another serious obstacle would have to be addressed.  The process of Wick rotation between real and imaginary time is relatively well controlled in theories with a static, eternal vacuum (such as the vacuum of a relativistic field theory).  In theories far from equilibrium, where the ``vacuum'' may not be eternal and static, the continuation would be much more subtle.  In the topological gravity of the Ricci flow, the saddle-point solutions to which the path integral localizes are the ``vacua'' of the theory, and they are often cosmologies with substantial time dependence, and even with singularities (recall Figs.~\ref{ffone} and~\ref{fftwo}).  They inherently represent systems very far from equilibrium, and one therefore would not expect that a simple analytic continuation interpolates between the real- and imaginary-time versions of the theory.  The full machinery of the Schwinger-Keldysh formalism for quantum systems far from equilibrium%
\footnote{For a recent discussion of the Schwinger-Keldysh formalism in the context of string theory, and for extensive references on the formalism, see \cite{neq}.}
may be needed in order to settle this intriguing question.  

\section{Summary and outlook}
\label{ssc}

In this paper, we have established contact between the mathematics of Ricci flow and topological quantum field theory.  It takes the form of a nonrelativistic topological quantum gravity, of the Lifshitz type.  

Even though this theory would perhaps be most interesting in $3+1$ dimensions, for most of the paper we presented our results in an arbitrary spatial dimension $D$.  This was possible primarily because we spent most of our work on constructing the action of the classical theory, with the correct gauge symmetries and BRST supersymmetry structure.  We expect the quantum properties of the theory to be more sensitive to $D$.  Note that the special case of $D=2$ would require some additional treatment already at the classical level, because of the well-known degeneracies that occur in Riemannian geometry in two spatial dimensions.   On the mathematical side, the $D=2$ analog of the Ricci flow is well-covered in the literature \cite{isenberg} (see also Ch.~5 of \cite{rfi}), and leads to a novel proof of the uniformization theorem for Riemann surfaces.  It should be possible to adjust the details of our construction to accommodate the special features of $2+1$ spacetime dimensions, which also happens to be the critical dimension in which quantum gravity of the Ricci flow is power-counting renormalizable.  

With the identification of Perelman's dilaton as our nonprojectable lapse function, and his $\CF$-functional as our superpotential, the localization equations in our topological quantum gravity represent a multi-parameter family of cousins to Perelman's original Ricci flow, parametrized by several coupling constants.  Yet, it might be difficult to see, in this forest of the many couplings in (\ref{eeforest}), where exactly the original Perelman Ricci flow equations are precisely reproduced.  In fact, since the localization equations (\ref{eeforest}) of the theory constructed in Section~\ref{sstime} are by design gauge invariant under foliation-preserving time reparametrizations --  a symmetry not shared by Perelman's equations --  they cannot reduce precisely to Perelman's flow equations for \textit{any} values of the couplings.  The precise embedding of Perelman's original flow into our theory requires a few additional steps, including a partial gauge fixing of our gauge symmetries, and we will present it in detail in our forthcoming paper \cite{prf}.    

One natural generalization that is accessible by our methods, but has not been discussed in the present paper, is the construction of topological gravity associated with the K\"ahler-Ricci flows, on spacetimes whose spatial slices $\Sigma$ carry a complex structure and whose dynamical spatial metric is K\"ahler.  This is an active area of current mathematical research, in particular in dimension $4+1$ (see \cite{song} or Ch.~2 of \cite{rf1}).  It would be very interesting to see what novel features the complex structure on space induces on the quantum gravity path integral, and the physical structure of the theory.

Another intriguing connection, not explored in the present paper, is the possible relation to quantum information theory.  In the mathematical context, Perelman's theory of the Ricci flow contains various quantities deservedly referred to as entropy.  In particular, the $\CF$-functional (and its close cousins the $\CW$- and $\CW_+$-functionals) belong to this category, and exhibit precise monotonicity properties, crucial for the proofs of various theorems about the behavior of the flow.  Their proper interpretation in the context of our topological quantum gravity is likely to be intimately connected to concepts of quantum information theory \cite{preskill,ewinf}, which have started playing a more dominant role in quantum field theory and quantum gravity in recent years.  

We fully expect that further study of topological quantum gravity associated with the Ricci flow should be beneficial both for physics and for mathematics: The wealth of mathematical results, generated especially in the past two decades, can teach us new lessons about quantum gravity, at least in the topological setting.  In turn, the methods of topological quantum field theory, which have proven so instrumental in influencing modern geometry in the past few decades, can now be extended to topological quantum gravity, and applied to the original mathematical theory of the Ricci flow.  In this context, it will be particularly interesting to study topological observables of the quantum theory.  While the BRST cohomology of our supermultiplets appears quite simple, and the ``moduli spaces'' of solutions are often elementary, it will be natural to probe the Ricci-flow spacetimes by \textit{extended} topological observables, such as topological strings and topological membranes.  Much of the mathematical ground for such observables has already been prepared, since extended spacetime probes of Perelman's flow have been studied extensively.  The mathematical results reviewed in \cite{tao} appear particularly promising, and suggest strongly that the topological quantum gravity introduced in this paper should naturally couple to topological brane excitations.  

\acknowledgments

One of us (P.H.) wishes to thank Kevin Grosvenor for illuminating discussions in the preliminary stages of this project.  The main results of this work have been presented by P.H. at the \textit{Marty Halpern Memorial Symposium} at Berkeley, California, in March 2019; and at \textit{FQMT 19: Frontiers of Quantum and Mesoscopic Thermodynamics} in Prague, Czech Republic, in July 2019.  P.H. thanks the organizers for the invitation and for creating a stimulating environment, and the conference participants for useful discussions.  This work has been supported by NSF grants PHY-1820912 and PHY-1521446.  

\appendix

\section{Prepotentials for the lapse and shift superfields}
\label{ssapp}

In order to gauge spatial diffeomorphisms and time reparametrizations, we introduced superfields  $N^i,S^i,\bar S^i$ and $E,\Theta,\bar\Theta$ respectively.  These superfields 
satisfy a complicated set of mutual constraints.  In order to make the superspace formulation simpler, especially in the quantum case, it would be beneficial to solve the constraints and express these constrained superfields in terms of unconstrained prepotential superfields.  The purpose of this Appendix is to identify such prepopotentials, both for the lapse and for the shift sector.

\subsection{Prepotential for the supervielbein}

Consider first the projectable Type B theory. Introduce an unconstrained projectable superfield $U(t,\theta,\bar\theta)$, the \textit{prepotential} for the projectable supervielbein.  $\Theta$ and $\bar\Theta$ are given by
\be
\Theta=-\frac{\RD U}{1+\dot U},\qquad \bar\Theta=-\frac{\bar \RD U}{1+\dot U}.
\ee
Such $\Theta$ and $\bar\Theta$ satisfy their nonlinear constraints.  $E$ then follows by plugging these expressions into the constraint that expresses $E$ in terms of $\Theta,\bar\Theta$ and their derivatives:
\be
E=\frac{1}{1+\dot U}.
\ee
It seems appropriate to refer to the prepotential $U$ of the lapse sector as ``prelapse.''

The extension to the nonprojectable Type B case is straightforward.  $U(t,\theta,\bar\theta,x^k)$ is now an unconstrained nonprojectable superfield, and
\be
\Theta=-\frac{\RD U-S^k\p_kU}{1+\dot U-N^j\p_jU},\qquad
\bar\Theta=-\frac{\bar \RD U-\bar S^k\p_kU}{1+\dot U-N^j\p_jU}.
\label{eethprel}
\ee
These expressions satisfy the full nonprojectable constraints (\ref{econ2}-\ref{econ4}), and give $E$ in terms of $U$.

The gauge transformations of the prepotential are
\be
\delta U=F+F\dot U+\Xi^k\p_kU,
\ee
and they correctly imply the standard gauge transformations for $\Theta,\bar\Theta$ and $E$.

Note that in (\ref{eethprel}), the constrained superfields $S^i, \bar S^i$ and $N^i$ of the shift sector appear explicitly.  In order to get an expression for the nonprojectable lapse superfieds in terms of only unconstrained superfields, we now have to find the prepotentials $V$ for the shift sector, express $S^i,\bar S^i$ and $N^i$ in terms of $V$, and substitute back in (\ref{eethprel}).

\subsection{Prepotential for the shift superfields}

Consider the shift superfields $N^i,S^i$ and $\bar S^i$ of Type B theory.  They can be expressed in terms of an unconstrained superfield prepotential $V^i$ as follows.  Denote by $\p V$ the matrix $\p_kV^i$, and by $\mathbb{I}$ the unit matrix $\delta_k^i$.  Write
\be
S^i=\RD V^k\left(\frac{1}{\mathbb{I}+\p V}\right)_{\! k}^{\ i},\qquad
\bar S^i=\bar \RD V^k\left(\frac{1}{\mathbb{I}+\p V}\right)_{\! k}^{\ i}.
\ee
These expressions again imply that the constraints \label{sconst} on $S^i$ and $\bar S^i$ are satisfied, and $N^i$ is then expressed in terms of $V^i$ via the constraints that gives $N^i$ in terms of $S^i,\bar S^i$ and their derivatives.  The vector prepotential transforms under the gauge symmetries as
\be
\delta V^i=\Xi^i+F\dot V^i+\Xi^k\p_kV^i.
\ee

While these expressions for the gauge superfields in terms of the prepotential superfields look quite simple, they are rather nonlocal and perhaps of limited practical use.  

\bibliographystyle{JHEP}
\bibliography{grf}
\end{document}